\documentclass[journal]{IEEEtran}
\ifCLASSINFOpdf
\else
\fi

\usepackage{graphicx}
\usepackage{amssymb,amsmath,latexsym,amsfonts}
\usepackage{epsfig}
\usepackage{comment}
\usepackage{color}
\usepackage[perpage,symbol]{footmisc}
\usepackage{algorithm}
\usepackage{algorithmic}
\usepackage{cite}

\newtheorem{definition}{Definition}
\newtheorem{theorem}{Theorem}

\newtheorem{corollary}{Corollary}
\def\done{\hspace*{\fill} \rule{1.8mm}{2.5mm}}

\newcommand{\SA}{ {\cal A} }

\newcommand{\SL}{ {\cal L} }
\newcommand{\SU}{ {\cal U} }
\newcommand{\SV}{ {\cal V} }
\newcommand{\SW}{ {\cal W} }
\newcommand{\SK}{ {\cal K}}

\newcommand{\SH}{ {\cal H} }

\newcommand {\bx}{\mbox{\boldmath $x$}}

\newcommand {\bv}{\mbox{\boldmath $v$}}

\begin{document}
%

\title{On The Robustness of Price-Anticipating Kelly Mechanism
\thanks{
Yuedong Xu, Tianyu Ni and Zhujun Xiao are with Research Center of Smart Networks and Systems, School of Information Science and Engineering, Fudan University; 
Jessie Hui Wang is with the Institute for Network Sciences and Cyberspace, Tsinghua University, Beijing 100084, China, and also with the Beijing National Research Center for Information Science and Technology, Beijing 100084, China; 
Xin Wang is with School of Computer Science, Fudan University; 
Eitan Altman is with Maestro Project, INRIA Sophia Antipolis, France. 
Email:ydxu@fudan.edu.cn, jessiewang@tsinghua.edu.cn, eitan.altman@inria.fr}}
\author{Yuedong Xu, Zhujun Xiao, Tianyu Ni, Jessie Hui Wang, Xin Wang and Eitan Altman}

\maketitle

\begin{abstract}


%
%


%

The price-anticipating Kelly mechanism (PAKM) is one of the most extensively used strategies to 
allocate divisible resources for strategic users in communication networks and computing systems. 
The users are deemed as selfish and also benign, each of which maximizes his individual utility 
of the allocated resources minus his payment to the network operator.  
However, in many applications a user 
can use his payment to reduce the utilities of his opponents, thus playing a misbehaving role. 
It remains mysterious to what extent the misbehaving user can damage or influence 
the performance of benign users and the network operator.

In this work, we formulate a non-cooperative game consisting of a finite amount of 
\emph{benign} users and one \emph{misbehaving} user. 
The maliciousness of this misbehaving user is captured by his willingness to pay to trade for unit 
degradation in the utilities of benign users. 
The network operator allocates resources to all the users via the price-anticipating Kelly mechanism. 
We present six important performance metrics with regard to 
the total utility and the total net utility of benign users, and the revenue of network operator under 
three different scenarios: with and without the misbehaving user, and the maximum. 
We quantify the robustness of PAKM against the misbehaving actions by deriving the upper and lower bounds of 
these metrics. With new approaches, all the theoretical bounds are applicable to 
an arbitrary population of benign users. 
Our study reveals two important insights: 
i) the performance bounds are very sensitive to the misbehaving user's willingness to pay 
at certain ranges; ii) the network operator acquires more revenues in the presence of the misbehaving user which might 
disincentivize his countermeasures against the misbehaving actions.
\end{abstract}


\IEEEdisplaynontitleabstractindextext
\IEEEpeerreviewmaketitle

\section{Introduction}
\label{sec:intrto}

Resource allocation is one of the fundamental issues in computer networks and computing systems 
that have received persistent studies from many aspects for decades. 
Taking our Internet as an example, the bandwidth is shared by heterogeneous 
population of users that vary in their types of traffic and their valuations to  
perceived network performance. A crucial question emerged, that is, how the resources should be 
shared among heterogeneous users efficiently. A bunch of early work proposed different economic approaches
to charge users according to their usage of network resources, thus addressing the heterogeneity 
in users and managing congestion in networks. Kelly in \cite{kelly,kelly2} proposed a market mechanism,
namely ``Kelly Mechanism'', in which each user submits a ``bid'' to the network operator, then 
the network operator determines the price of each link after collecting all the bids and allocates 
the resources to each user in proportion to his bid. 
Kelly mechanism consists of two scenarios: price-taking users and 
price-anticipating users. In the former, a price-taking user bids to the network operator with 
the observation of the per-unit of bandwidth price. In the latter, a price-anticipating user is strategic in
the sense that he takes into consideration how the per-unit of bandwidth price is influenced by his bid. 
Given the bidding vector of all the strategic users, he knows that the resources allocated 
to him are proportional to his bid, but inversely proportional to the sum of all bids. 
The higher bid a user pays, the more resources will be allocated to him and less to the others. 
Thus, a game is introduced because the bid of each user affects the payoffs of other users, and it is 
suitable to capture the competition for limited and divisible bandwidth resources \cite{yang2007vcg}. 

When users are price-anticipating, the system utility at Nash Equilibrium (NE) will deviate from that at  
social optimum. Johari and Tsitsiklis in \cite{johari} have proved that the price-anticipating Kelly mechanism (PAKM) 
yields at most a 25\% efficiency loss at the equilibrium. This optimistic result is deemed as the cornerstone of PAKM, which 
has rarely been re-examined in the following decade. Owing to its simplicity and guaranteed performance, 
the price-anticipating Kelly mechanism has gained popularity beyond 
the bandwidth allocation of fixed-line \cite{yang2007vcg,massoulie1999,yang2013price} and 
mobile networks \cite{akkarajitsakul2009,caballero2016multi}, 
which we name a few new applications as below. 
 \begin{itemize}

\item \textit{Server Cluster/5G Network Slicing:} Feldman \emph{et al.} adopted the price-anticipating Kelly mechanism 
to allocate computing resource to selfish users with a fixed budget in distributed clusters \cite{feldman2005price}. Caballero \emph{et al.} 
applied this mechanism to perform network slicing in 5G mobile networks, 
i.e. dividing the mobile network infrastructure into logical networks \cite{caballero2016multi,caballero2017}. 

\item \textit{Crowdsourcing Incentivization:} Yang \emph{et al.} utilized PAKM to incentivize 
mobile users to contribute their sensing time in crowsensing applications \cite{yangmobicom}. Ghosh and McAfee analyzed 
the economics of incentivizing high-quality user generated content using PAKM \cite{ghosh2012crowdsourcing,ghosh2011incentivizing,luo2015crowdsourcing}.

\item \textit{Visibility/Advertising Competition Online:} Altman \emph{et al.} modeled the competition for users' 
attention in the timeline of online social networks (OSNs) \cite{Altman1,Altman2,Altman3,Altman4,Altman5} in which the total attention is 
divided by multiple owners according to PAKM. Bimpikis and Ozdaglar \emph{et al.} \cite{Ozdaglar} studied  
the optimal targeted advertising in OSNs where 
the competition of two brands can take the form of PAKM. 

\item \textit{Blockchain Mining:} The probability that a bitcoin miner first discovers a block is 
determined by PAKM where the bid of the miner is his Hashrate to solve cryptographic puzzles \cite{blockchain}.



\end{itemize}
 
Though fruitful, previous work laid the foundation on the \emph{selfishness} (or benignancy interchangeably) assumption that each user aims to 
maximize his payoff, and this payoff reflects his valuation or satisfaction to the share of resource. 
Beyond selfishness, a user can be misbehaving in which his goal is to reduce the utilities of all other selfish users. 
There have been a number of real-world counterparts, taking different forms in different applications. 
For instance, an attacker can hijack zombie computers so as to generate a high volume of misbehaving traffic to 
perform denial-of-service (DoS) attacks. Authors in \cite{Vulimiri} presented PAKM to 
allocate bandwidth among the benign users and the attacker. For visibility competition in OSNs, 
the fraction of viewers' attention is determined by PAKM \cite{Altman1} 
where the misbehaving advertiser can decrease the visibility of the benign advertisers  
by posting excessive messages to social media, thus flushing down the relatively old messages of benign ones. 
They can even pay a fee to the network operator against benign advertisers so that the messages of misbehaving users 
can be displayed in the timelines of potential viewers for a certain period. Furthermore, the price-anticipating 
Kelly mechanism has been applied to social security models 
\cite{warneryd2003information,rietzke2013robustness,bevia2010peace} that can also sharpen the understanding of 
adversary competition over networks. 
Therefore, it is necessary to figure out the damage caused by the misbehaving user rigorously. 
In the light of the heterogeneity of benign and 
misbehaving users, case-by-case studies neither capture the efficiency loss in the general settings, nor 
provide insights on the countermeasures against the misbehaviors.

 In this paper, we model the competition of users as a non-cooperative game where the players
consist of a finite number of benign users and one misbehaving
user. The strategy of a player is his bid submitted to the network operator. 
The utility of a benign user is his satisfaction on the allocated resources, and the net utility 
is defined as his utility minus the bid paid to the network operator. 
While the satisfaction of the misbehaving user is not determined by the resourced allocated by him, but 
the losses that the benign users suffer from the misbehaviors. 
Similar to \cite{Vulimiri}, the hostility of the misbehaving user is captured by his \emph{willingness to pay}  
or \emph{willingness factor}. The willingness factor refers to a scalar that 
the misbehaving user wishes to use a unit cost or payment to trade for the loss of aggregate utility of the 
benign users. Hence, his objective is to maximize the utility loss of the benign users minus his 
payment to the network operator.

Our work distinguishes from the literature in two aspects: \emph{new performance metrics} and \emph{new analytical approaches}. 
When the misbehaving user performs actions, we are faced with the following fundamental question: 
\emph{To what extent the misbehaving user can damage or influence the performance of the benign users and the 
network operator at a Nash Equilibrium?} 
In \cite{Vulimiri}, authors present two metrics, $\SU$ and $\SL$.  
Here, $\SU$ denotes the total utility of the benign users, reflecting the extent that all the 
benign users enjoy their allocated resources. $\SL$ 
is the sum of $\SU$ and the misbehaving user's bid submitted to the network operator. 
In other words, the benign agents and the network operator are treated as ``good'' players 
so that $\SL$ is their total profit. 
However, using $\SL$ as a metric conceals the damage on the net utility 
of benign users, and the impact on the revenue of the network operator that they should be scrutinized separately. 
Both the benign users and the network operator are ``good'' but playing different roles and possessing different objectives. 
On one hand, each benign user eventually cares about his net utility of the allocated flow rate or the online visibility etc.
On the other hand, the revenue harvested by the network operator may influence his countermeasures against the 
misbehaving user. Hence, we raise two new metrics, the total net utility of benign users $\SV$ and 
the total revenue of the network operator $\SW$. To quantify the efficiency loss in the presence of 
the misbehaving user, we define three representative scenarios:
i) NE of the game excluding the misbehaving user ($nom$), ii) NE of the game consisting of the
misbehaving user ($mal$), 
iii) the maximum measures excluding the misbehaving user ($max$).

We analyze the above performance metrics in different scenarios via their upper and lower bounds. 
The ratios $\frac{\SU_{mal}}{\SU_{max}}$, $\frac{\SV_{mal}}{\SV_{max}}$ and 
$\frac{\SW_{mal}}{\SW_{max}}$ compare the metrics in the presence of the misbehaving user 
with the maximum metrics excluding him. 
The ratios $\frac{\SU_{mal}}{\SU_{nom}}$, $\frac{\SV_{mal}}{\SV_{nom}}$ and 
$\frac{\SW_{mal}}{\SW_{nom}}$ demonstrate the changes of metrics 
between the two NEs excluding and including the misbehaving user. 
Compared with previous studies \cite{johari} and \cite{Vulimiri}, our work considers two important 
extra metrics ($\SV$ and $\SW$), derives both the 
lower and the (approximate) upper bounds of these ratios 
instead of the lower bounds only. The theoretical results 
are obtained for any number of benign users other than the regime of the infinite number of benign users.  
The approaches of analyzing the bounds are also different from those in \cite{johari} and \cite{Vulimiri}, and 
they are far beyond simple calculations because the NEs at different games involve different sets of players. 
The lower bounds manifest how the utility and the net utility of benign users, and the revenue of network operator are influenced by the maliciousness of the misbehaving user in the worst case. 
The upper bounds are used to gauge how good the performance metrics are at the NE 
by showing their gaps to the best achievable values. 
Our study reveals three interesting phenomena: i) the willingness factor of the misbehaving user 
in certain ranges can remarkably alter the upper and the lower 
bounds of the benign users' total utility and 
total net utility; ii) the network operator obtains a better revenue from the
misbehaving actions that may undermine his will to take them down; 
iii) the lower bounds are improved when the network operator charges the misbehaving 
user a higher price, and the lower bound of total utility over maximum may be worse off when 
a benign user is mischarged a higher price.

We summarize our main results and contributions as follows.

\begin{itemize}

\item We present new metrics to quantify the robustness of price-anticipating Kelly mechanism in the 
presence of a misbehaving user. 
	
\item For general utility functions of benign users, we quantify the lower bounds of $\frac{\SU_{mal}}{\SU_{max}}$, $\frac{\SU_{mal}}{\SU_{nom}}$, $\frac{\SV_{mal}}{\SV_{max}}$ and $\frac{\SV_{mal}}{\SV_{nom}}$
	that are jointly determined by the number of benign users ($N$) and the willingness factor ($\theta$). 
	Their upper bounds are also investigated when the utility functions are linear to the allocated resources.


\item We compute the upper and lower bounds of $\frac{\SW_{mal}}{\SW_{max}}$ and 
	$\frac{\SW_{mal}}{\SW_{nom}}$ for linear utility functions. As $\theta$ grows, the both bounds 
	of $\frac{\SW_{mal}}{\SW_{max}}$ increase. The ratio $\frac{\SW_{mal}}{\SW_{nom}}$ is no worse than 1 and its upper bound is $\infty$.

\item When price differentiation is taken as a countermeasure, charging a higher 
price on the per unit of resource of the misbehaving (resp. benign) user may improve (resp. lower down) the performance metrics as well as lower bounds, but the impact on the revenue of network operator is uncertain.
	
\end{itemize}

The remainder of this paper is organized as follows. Section \ref{sec:model} presents the mathematical 
model of the price-anticipating Kelly mechanism with a misbehaving user. 
Section \ref{sec:bounds} derives the upper and the lower bounds of the benign users in PAKM. 
We validate those bounds numerically in Section \ref{sec:simu}. Section \ref{sec:implications} 
analyzes the impact of different price differentiation schemes on the performance metrics. 
Section \ref{sec:related} describes related work 
and Section \ref{sec:conclusion} concludes this work.

\section{Model and Measures}
\label{sec:model}

In this section, we construct the basic game model of price-anticipating Kelly mechanism with selfish as well as misbehaving users, 
and present a set of new performance metrics to quantify the impact of misbehaving users on the user utility.

\subsection{Basic Model}

We suppose that there exist $N{+}1$ users denoted by $\SA=\{\SA_0,\SA_1,{\cdots}, \SA_N\}$. 
These users are divided into two groups where those in the set 
$\{\SA_1,\cdots,\SA_{N}\}$ are \emph{benign} and $\SA_0$ is \emph{misbehaving}. 
Every user, either benign or misbehaving, pays a certain bid to acquire a fraction of network 
resources. The purpose of the benign user is to achieve an optimal tradeoff between his 
utility on his obtained resources and the payment to the network operator. 
On the contrary, the misbehaving user aims to reduce the utility of all the benign users, 
considering his payment. Throughout this work, we assume that the total amount of resources 
is normalized as 1. Note that all the results hold for any finite amount of resources.

{\bf Utility Model:}
We hereby describe the utility functions of all the users. Define $\bx=\{x_0, x_1,\cdots,x_N\}$ as the vector of bids paid to the network operator and define 
$z:=\sum_{i=0}^{N}x_i$ as the total amount of bids. The fraction of 
$\SA_i$'s bid over the total is denoted as
\begin{eqnarray}
	d_i = \frac{x_i}{z} = \frac{x_i}{\sum_{j=0}^{N}x_j}
	\label{eq:d_i}
\end{eqnarray}
that is also the fraction of resources allocated to $\SA_i$\footnote{When the resource is not divisible, e.g. a time slot or a channel in wireless networks, 
	or an advertising location in webpages, $d_i$ refers to the probability of the $i^{th}$ user to acquire this resource. }.
Let $\mathbf{d}{=}\{d_0,d_1,{\cdots}, d_N\}$ be the vector of all the fractions. 
Denote by $U_{\SA_i}(d_i)$ the utility of $\SA_i$ of utilizing the fraction of resource $d_i$. 
The net utility of a benign user $\SA_i$, $V_{\SA_i}(\bx)$, 
is defined as the difference between the utility and the bid paid to the network operator,
\begin{eqnarray}
V_{\SA_i}(\bx) = U_{\SA_i}(d_i(\bx))  - x_i, \quad \forall 1\leq i \leq N. 
\label{eq:model_no1}
\end{eqnarray}

	The objective of the misbehaving user is to reduce the utilities of all other benign users. 
	Hence, the 
	utility and the net utility of  $\SA_0$ are expressed as
	\begin{eqnarray}
	\SU_{\SA_0}(\bx) \!\!&=&\!\! - \theta \sum\nolimits_{i{=}1}^{N}U_{\SA_i}(d_i(\bx)) \\
	V_{\SA_0}(\bx) \!\!&=&\!\! - \theta \sum\nolimits_{i{=}1}^{N}U_{\SA_i}(d_i(\bx)) - x_0,
	\label{eq:model_no2}
	\end{eqnarray}
	where $\theta>0$ is the \textit{willingness to pay} (or \emph{willingness factor}) of the misbehaving user. 
This means that the misbehaving user would like to use $\theta$ dollars to trade for a unit 
reduction of benign users' utility \cite{Vulimiri}. 
Note that $\theta$ captures $\SA_0$'s resolution to reduce the total utilities of benign users. 
At an extreme point $\theta=0$, $\SA_0$ does not participate in the timeline competition in any situation, 
while at the other extreme point $\theta\rightarrow \infty$, $\SA_0$ tries all means to perform 
DDoS attack against benign users, or to create messages with an infinite intensity to flush away those 
of benign users in targeted online advertising. 
We adopt the same utility model as that in \cite{Vulimiri}, but the major differences 
lie in the analysis of novel performance metrics that shed light on the impact of the misbehaving user on the 
efficiency loss of the system. 

Similar to the prior work on network 
resource allocation \cite{johari} and \cite{Vulimiri}, we employ the following 
assumption on the utility of benign users. Note that the utility of a benign user is a function of its allocation, and also a function of the scalar of the bids indirectly. We hereby clarify the difference and the 
relevancy of hypotheses made on each of them.

\noindent\textit{Assumption \textbf{P}:} The utility of the $i^{th}$ benign user, $U_{\SA_i}(d_i(\bx)): \mathbb{R}_{+}^N\mapsto \mathbb{R}_{+}$ satisfies:

\begin{itemize}
	
	\item (\emph{P1: Monotonicity and Concavity}) non-negative, strictly increasing, concave and continuously differentiable in the allocation $d_i$ for all $i\in [1, N]$;

	\item (\emph{P2: Monotonicity and Convexity}) non-negative, strictly decreasing and strictly convex in $x_0$.  
	
\end{itemize}

In the first part \emph{P1}, a benign user acquires a better utility if his allocation is larger. However, further increasing his 
share of resources does not yield an increasing marginal utility. Such a utility function can be interpreted as the satisfaction of the
benign user to the bandwidth in communication networks, or the visibility in online media. As a direct deduction of \emph{P1}, 
$U_{\SA_i}(d_i(\bx))$ is a strictly increasing and strictly concave with regard to $x_i$ (even if $U_{\SA_i}(d_i(\bx))$ is a linear function of 
$d_i(\bx)$). The benign user $\SA_i$ can obtain a higher utility if he bids a higher $x_i$, yet further increasing 
his bid yields a shrinking marginal utility. 
Meanwhile, \emph{P1} coincides with the assumption of \emph{Theorem 2} in \cite{johari} 
and the assumption of \cite{hajek2} that guarantee the existence and 
uniqueness of NE without considering the misbehaving user.

In the second 
part \emph{P2}, the utility of each benign user is a strictly decreasing and strictly convex function of $x_0$. Hence, 
the utility of the misbehaving user $\SU_{\SA_0}(\bx)$ is strictly increasing and strictly concave with regard to $x_0$.
When the misbehaving user 
pays a higher bid $x_0$, the utility of each benign user decreases, and further increasing $x_0$ causes the moderative 
reduction of the utilities of all the benign users. 

We further define a special type of utility function satisfying assumption \textbf{P}:
\begin{definition} \textit{(Linear Utility Function)} 
	The utility of benign user $\SA_i$ is a linear function if it is in the form 
	\begin{eqnarray}
	U_{\SA_i}(d_i) = v_id_i, \quad \forall i=1,\cdots, N,
	\label{eq:linearutility}
	\end{eqnarray}
	where $v_i$ is regarded as $\SA_i$'s valuation or satisfaction to the allocated resources. The valuations
	are sorted in a decreasing order, i.e. $v_1 {\geq} v_2 {\geq} \cdots {\geq}v_{N}$ with $v_1$ normalized as 1. 
	\label{def:linearutility}
\end{definition}

Other widely used utility functions \cite{kelly2}\cite{alpha} that satisfy assumption \textbf{P} include the log-utility function and alpha-fair utility function. They take the following forms respectively:
\begin{eqnarray}
	U_{\SA_i}(d_i) = v_i\log(1+d_i) \;\;\; \text{and} \;\;\; U_{\SA_i}(d_i) = v_i\frac{d_i^{1-\alpha}}{1-\alpha}, \;\;\; \forall i\in[1, N], \nonumber
\end{eqnarray}
where $\alpha$ is a constant in the range $(0, 1)$.

{\bf Game Model:} We formulate the competition of all the benign and misbehaving users as a noncooperative game 
denoted by $\mathbf{G}$ that comprises three key elements:

\noindent\textit{- Players:} a set of users $\{\SA_0, \SA_1,\cdots, \SA_N\}$;

\noindent\textit{- Strategies:} each player's strategy is the bid paid to the network operator, i.e. $x_i$ for all $i=0,1,\cdots,N$;

\noindent\textit{- Payoffs:} the payoff of a player is his net utility, i.e. $(V_{\SA_0}, V_{\SA_1}, \cdots, V_{\SA_N})$.

For the game $\mathbf{G}$, we define its Nash Equilibrium (NE) as the following.
\begin{definition} \textit{(Nash Equilibrium)} 
	Let $\bx=(x_0, x_1,\cdots,x_N)$ be the vector of bids paid to the network operator, and 
	$\bx_{-i}{=}\bx{\setminus}\{x_i\}$. A strategy profile $\bx^*$ is a Nash equilibrium if 
	$V_{\SA_i}(x_i^*, \bx_{-i}^*) {\geq} V_{\SA_i}(x_i, \bx_{-i}^*)$ for any $x_i \neq x_i^*$ and any $\SA_i\in\SA$.
	\label{def:ne}
\end{definition}

\subsection{Performance Measures}

The competitions take place among the benign users, and  
between the benign users and the misbehaving user. 
It is expected that the misbehaving user will reduce the utility and the net utility of the benign users, and 
affect the revenue of the network operator. 
We define three measures to quantify the outcomes of competition: \emph{total utility} ($\SU$) 
and \emph{total net utility} ($\SV$) of benign users as well as \emph{total network operator's revenue} ($\SW$). 
Formally, there have
\begin{eqnarray}
\SU = \sum\nolimits_{i=1}^{N}U_{\SA_i}, \quad \SV =  \SU -  \sum\nolimits_{i=1}^{N} x_i, \quad \SW = \sum\nolimits_{i=0}^{N}  x_i  \nonumber.
\end{eqnarray}
Here, $\SU$ reflects the total satisfaction of benign users with respect to their allocated resources. 
In addition, $\SU$ is equivalent to the sum of all the net utilities of benign users and their bids acquired by the network operator. 
Johari and Tsitsiklis in \cite{johari} proved that the competition among the benign users leads to 
at most 25\% loss of $\SU$ compared with the social optimum. 
Vulimiri and Agha \emph{et al.} generalized the analysis of $\SU$ by introducing a misbehaving user \cite{Vulimiri}. 
However, taking $\SU$ as the only performance measure may overlook the different roles that the benign users 
and the network operator play in many applications. 
For instance, in the DDoS attack, each benign user cares about his loss of net utility caused by the attacker, 
and the network operator is interested in the total price paid by all the users using the bandwidth.  
In the targeted online advertising, it is also important to examine the efficiency loss of net utilities obtained by the 
benign users, and the total revenue received by the OSN. 
Therefore, we propose two new metrics, $\SV$ and $\SW$, to analyze the net utility of benign users and the 
revenue of the network operator in the presence of a misbehaving user. 

To capture the efficiency of the misbehaving user to neutralize the gains of 
the benign users and his impact on the network operator's revenue, we compare $\SU$ (resp. $\SV$ and $\SW$) 
in three scenarios: $\mathbf{MAL}$, $\mathbf{NOM}$ and $\mathbf{MAX}$. 

\begin{itemize}

\item $\mathbf{MAL}$ refers to the game \textbf{consisting} of all the players in $\SA$; 

\item $\mathbf{NOM}$ refers to the game \textbf{excluding} the misbehaving user 
$\SA_0$;

\item $\mathbf{MAX}$ refers to the maximum achievable metrics when \textbf{excluding} the misbehaving user $\SA_0$.

\end{itemize}

Note that in each of $\mathbf{MAL}$ and $\mathbf{NOM}$, there 
	exists a unique Nash equilibrium that will be proved in section \ref{subsec:ne}. 
	Before diving into the mathematical analysis, we introduce the performance metrics first.
In general, the utility functions of the benign users are rather diverse; 
one has to investigate enormous cases so to understand the impact of misbehaving behavior 
on the above metrics. 
Instead of pursuing case-by-case studies, we resort to the worst and the best 
performance for versatile utility functions. 
Therefore, our primary goals are to quantify the following bounds:

\begin{itemize}
	\item \textbf{B1}: $\frac{\SU_{mal}}{\SU_{max}}$, the gap between 
	the total utility (or satisfaction) at the NE of $\mathbf{MAL}$ and the 
	maximum total utility (or satisfaction) of benign users.	
	
	\item \textbf{B2}: $\frac{\SU_{mal}}{\SU_{nom}}$, the damage to the 
	total utility (or satisfaction) of benign users at two NEs of $\mathbf{MAL}$
	and $\mathbf{NOM}$ respectively.

	\item \textbf{B3}: $\frac{\SV_{mal}}{\SV_{max}}$, the dissipation of 
	total net utility of benign users at the NE of $\mathbf{MAL}$ in comparison to the 
	maximum total net utility of benign users. 
	
	\item \textbf{B4}: $\frac{\SV_{mal}}{\SV_{nom}}$, the damage to the total net utility 
	of benign users at two NEs of $\mathbf{MAL}$ and $\mathbf{NOM}$ respectively. 
	
	\item \textbf{B5}: $\frac{\SW_{mal}}{\SW_{max}}$, the ratio between the network operator's revenue at the 
	NE of $\mathbf{MAL}$ and the \emph{maximum} revenue of the network operator in $\mathbf{MAX}$. 
	
	\item \textbf{B6}: $\frac{\SW_{mal}}{\SW_{nom}}$, the ratio between the 
	network operator's revenue at the NE of $\mathbf{MAL}$ and that at the NE of $\mathbf{NOM}$. 
\end{itemize}

The bounds \textbf{B1}, \textbf{B3} and \textbf{B5} mainly capture the absolute performance at the NE 
with the possible emergence of the misbehaving user (compared with the 
maximum performance free of the misbehaving 
user). While the bounds \textbf{B2}, \textbf{B4} and \textbf{B6} look into the relative changes 
brought by the possible emergence of the misbehaving user.

In \cite{Vulimiri}, the total utility of benign users 
is the main concern. The lower bound of $\frac{\SU_{mal}}{\SU_{max}}$ 
has been analyzed in the extreme with unlimited number of players. 
In comparison, the major differences between \cite{Vulimiri} and our work are listed below:

\textbf{i)}: We consider a finite number of benign users where the bounds of \textbf{B1} in 
\cite{johari} and \cite{Vulimiri} are special cases of ours at the regime of infinite number of users;
		
\textbf{ii)}: The bounds of \textbf{B2}$\sim$\textbf{B6} have not been studied previously;
	
\textbf{iii)}: We further obtain the upper bounds
	to exhibit the best performance measures for benchmark linear utility functions;
	
\textbf{iv)}: We study the bounds of the network operator's revenue;
	
\textbf{v)}: We analyze the impact of price differentiation as a countermeasure on the utilities 
of different economic entities and their bounds.

\section{Efficiency Loss of Benign users}
\label{sec:bounds}

In this section, we obtain the bounds of performance measures for 
the price-anticipating Kelly mechanism with a misbehaving user. Our analyses provide 
deep understandings on how (in)efficient the misbehaving user can reduce 
the benefits of the benign users. 

\subsection{Nash Equilibrium}
\label{subsec:ne}

We first echo that the game $\mathbf{G}$ induces a unique Nash equilibrium. 
The uniqueness has been proved in Kelly's mechanism (price anticipating bandwidth sharing game) 
with only benign users \cite{johari}, and in the security game \cite{Vulimiri}. 
Here, a couple of known results are summarized. 
\begin{theorem}\textit{(Uniqueness of Nash Equilibrium)\cite{Vulimiri}}
	\label{theorem:no1_uniqueness}
	The game $\mathbf{G}$ has a unique Nash Equilibrium $\bx^*{\geq} 0$ under assumption \textbf{P}, where
	at least two components of $\bx^*$ are positive. 
	\label{theorem:uniqueness}
\end{theorem}
\noindent \textit{Remark 1:} 
With assumption \textbf{P}, the utilities of 
all the players including the benign and the misbehaving users are concave 
functions of their individual bids, thus constituting a \emph{concave game}
\cite{rosen}.

The above theorem also implies that a subset of users may not pay a fee to the network operator. 
We claim that $\SA_i$ does not \textbf{\emph{participate}} in the competition if his payment has $x_i^*=0$. 
In the standard Kelly's mechanism, at least two players participate at the NE. 
This is also true in the presence of the misbehaving user with the \emph{specialty} that these two players can be 
one benign user and the misbehaving user. Let us first examine 
the participation of players at the NE. 
To find the NE of this continuous game, we take the first-order partial 
derivative 
of $U_{\SA_i}(\bx)$ over $x_i$, $\partial U_{\SA_i}(\bx)/\partial 
	x_i$, for all $i\in[0, N]$. The bid of $\SA_i$ at the NE is the best 
	response of the bids of his opponents. Therefore, $\partial U_{\SA_i}(\bx)/\partial x_i$ is 0 if the best response $x_i^*> 0$ and 
	becomes non-positive if $x_i^*$ is 0 (i.e. at the boundary). The conditions of the NE 
	(denoted as \textbf{NE\_CONDs}) are summarized below (recalling $z=\sum_{i=0}^{N}x_i$): 
\begin{eqnarray}
&&U_{\SA_i}'(d_i^*)\frac{z_{-i}^*}{(z^*)^2} -1  \left\{\begin{matrix}
= 0   &\textrm{ if }  x_i^* > 0\\
\leq 0 &\textrm{ if }  x_i^* = 0 
\end{matrix}\right. \quad \forall i\geq 1,
\label{eq:onetarget_opt1}\\
&&\theta \sum\nolimits_{i{=}1}^{N} U_{\SA_i}'(d_i^*) \frac{x_{i}^*}{(z^*)^2} - 1\left\{\begin{matrix}
=0   &\textrm{ if }  x_0^* > 0 \\
\leq 0 &\textrm{ if }   x_0^* = 0 
\end{matrix}\right.
\label{eq:onetarget_opt2}
\end{eqnarray}
where $U_{\SA_i}'(d_i)$ is the derivative of $U_{\SA_i}(d_i)$ over $d_i$. 
\textbf{NE\_CONDs} in Eqs.\eqref{eq:onetarget_opt1} and \eqref{eq:onetarget_opt2} 
do not yield close-form expression of the NE, thus not allowing further understanding on its property in general.

We compute the NE when the utilities of the 
benign users are linear functions of the allocated resources. 
Our purpose is to scrutinize the participation of players on top of an explicit-form NE. The linear utility function is crucial to the analysis of 
theoretic bounds later on.

\begin{theorem} \textit{(Nash Equilibrium for Linear Utility)} The NE with linear utility functions $\{U_{\SA_i}(d_i)\}_{i=1}^N$ satisfies:
	\begin{itemize}
		\item if $n$ out of $N$ benign users participate and there exists 
		\begin{eqnarray}
		\theta <\frac{n-1}{\sum\nolimits_{j=1}^{n}v_j\sum\nolimits_{j=1}^{n}1/v_j {-}n(n{-}1)},
		\end{eqnarray}
		the NE strategy is computed by
\begin{eqnarray}
\left\{\begin{matrix}
x_i^* = \big(\frac{\sum\nolimits_{j=1}^{n}1/v_j}{n{-}1} {-} \frac{1}{v_i}
\big)\big(\frac{n{-}1}{\sum\nolimits_{j=1}^{n}1/v_j}\big)^2  \\
x_0^*=0  
\end{matrix}\right.;
\label{eq:linearutility_no1}
\end{eqnarray}

       \item  if otherwise, the NE strategy is computed by
\begin{eqnarray}
\left\{\begin{matrix}
x_i^* = \big(\frac{n+1/\theta}{\sum\nolimits_{j=1}^{n}v_j} {-} \frac{1}{v_i}
\big)\big(\frac{\sum\nolimits_{j=1}^{n}v_j}{n{+}1/\theta}\big)^2 \\
x_0^*=\frac{\sum\nolimits_{j=1}^{n}v_j}{n+1/\theta}
\big( \frac{\sum\nolimits_{j=1}^{n}v_j\sum\nolimits_{j=1}^{n}1/v_j}{n+1/\theta}{+} 1 {-} n\big)
\end{matrix}\right.;
\label{eq:linearutility_no2}
\end{eqnarray}

		\item the number of benign users with positive bids, $n^*$, is searched 
		from $N$ to $1$ until $x_i^* \ge 0$ for $1\leq i\leq n$;
		
		\item the utility and the net utility of benign users are decreasing functions of $\theta$ upon $x_0^* > 0$.
	\end{itemize}
	\label{theorem:linear_ne}
\end{theorem}

\noindent\textbf{Proof:} Please refer to the appendix. \done

A simple search method is shown in Algorithm \ref{algo1}.
The first step is to compute $\bx^*$ by assuming the participation 
of $n=N$ benign users at the NE.  
If $x_i^*$ is positive for all $\SA_i\in\SA$, the NE is obtained.
Otherwise, if any $x_i^*$ is negative, this means that some of the 
benign users or the misbehaving user do not
participate in the competition. By removing the concurrent benign user
with the smallest valuation or the misbehaving user, 
we proceed to search until $n{=}1$.


\begin{algorithm}
\caption{: Searching the NE for Linear Utility Functions}
\label{power_alc}
\begin{algorithmic}[1]
\renewcommand{\algorithmicrequire}{\textbf{Input: $\mathbf{v}$, $\theta$, $N$}; \quad \textbf{Output:} $\mathbf{x^{\star}}$}
\REQUIRE ~~ \\
\STATE \textbf{sort} $v_i$ in the descending order
\STATE  \textbf{for} $n = N$ to $1$ 
\STATE \ \ Compute $x_i^*$ using Eq.\eqref{eq:linearutility_no2}
\STATE \ \ \textbf{if} $x_i^{*} \geq 0$ for all $i=0,\cdots, n$
\STATE \ \ \ \ \textbf{exit}
\STATE \ \  \textbf{else}
\STATE \ \ \ \ Compute $x_i^*$ using Eq.\eqref{eq:linearutility_no1}
\STATE \ \ \ \ Compute the NE conditions in \eqref{eq:onetarget_opt1} and \eqref{eq:onetarget_opt2} 
\STATE \ \ \ \ \textbf{if} $\theta \sum\nolimits_{i{=}1}^{N} U_{\SA_i}'(d_i^*) \frac{x_{i}^*}{(z^*)^2} \leq 1$ and $x_n^* \geq 0$
\STATE \ \ \ \ \ \ \textbf{exit}
\STATE \ \ \ \ \textbf{end}
\STATE \ \ \textbf{end}
\STATE \textbf{end}
\end{algorithmic}
\label{algo1}
\end{algorithm}

We further present general properties w.r.t. the participation of benign 
users when the utilities are linear to their fractions of acquired resources. 

\begin{theorem} \textit{(Participation of Players for Linear Utility)}
	The NE $\bx^*$ has the following properties:
	\begin{itemize}
		\item if $\theta > \frac{N{-}1}{N}$, then there always has $x_0^* > 0$;
		
		\item if $v_i>v_j$ and $x_j^*>0$, then $x_i^* {>} x_j^* {>} 0 $ $\forall 1\leq i,j \leq N$;
		
		\item if $v_i=v_j$ for all $i,j$, then $x_i^* > 0$ for all $1\leq i \leq N$. 
	\end{itemize}
	\label{theorem:no2_participation}
	\vspace{-0.1cm}
\end{theorem}
\noindent\textbf{Proof:} Please refer to the appendix. \done

Theorem \ref{theorem:no2_participation} shows the sufficient conditions that the benign users 
or the misbehaving user participate in the competition.
When $\theta$ is larger than a fixed threshold, $\SA_0$ will always play against the benign users, 
regardless of the valuations of benign users (with the normalization of $v_1$).
The second bullet shows that the benign users with higher valuations are more likely to
participate at the NE. 

\subsection{Lower and Upper Bounds for Linear Utility Functions}

The NE of a price-anticipating Kelly mechanism is easy to solve, and 
it is determined by the vector of valuations and the willingness factor in 
the linear case. The valuations are chosen arbitrarily such that 
an important question is concealed: \emph{how harmful is the misbehaving user's 
	strategy on the utilities of benign users?} 
Trying as more examples of $\bv$ as possible numerically may give rise to 
the bounds of performance metrics for certain parameters or certain utility function, 
however, it fails to deliver the fundamental qualitative properties. 
To this goal, we compare the NEs in three scenarios: $\mathbf{MAL}$, $\mathbf{NOM}$ and $\mathbf{MAX}$. The metrics that are studied consist of 
the utility of benign users, the net utility of benign users and the 
revenue of network operator. 
The following theorem states the lower bounds of Kelly mechanism with 
linear utility functions.

\begin{theorem} \textit{(Lower Bounds)} 
For linear utility functions of benign users, the lower bounds of \textbf{B1}$\sim$\textbf{B6} are as follows.
\begin{itemize}
	\item \textit{\textbf{B1:} (NE Utility over Maximum)} 
	\begin{eqnarray}
	\frac{\SU_{mal}}{\SU_{max}} \geq \left\{\begin{matrix}
	\;(1{+}\theta)^{{-}1} ,\;\; \textrm{ if } \;\; 
	\theta {>} \big((\sqrt{\frac{N}{N{-}1}}{+}1)^2 {-} 1\big)^{{-}1} \\
	\; 1{-}(N{-}1)(\sqrt{N}{-}\sqrt{N{-}1})^2, \;  \textrm{ otherwise }
	\end{matrix}\right.
	\label{eq:theorem_lowerbounds_no1} 
	\end{eqnarray}
	\noindent where the bound is tight with $v_i<\frac{\theta}{1+\theta}$ for $i\geq 2$ in the upper formula, 
	and is tight with $v_i = \sqrt{N(N{-}1)}{-}(N{-}1)$ 
	for $i\geq 2$ in the lower formula. 
		As $N\rightarrow \infty$, the asymptotic lower bound is 
		\begin{eqnarray}
		\frac{\SU_{mal}}{\SU_{max}} \geq \min\{\frac{3}{4}, \frac{1}{1{+}\theta}\}
		\label{eq:theorem_lowerbounds_no2}
		\end{eqnarray}
		where $3/4$ is reached at $v_i=1/2$ for $i\geq 2$ \footnote{The asymptotic lower bound  
			of \textbf{B1} for $N\rightarrow \infty$ has been proved in \cite{Vulimiri} using a different approach.}.

	\item \textit{\textbf{B2:} (NE Utility With/Without Misbehaving User)} 
	\begin{eqnarray}
	\frac{\SU_{mal}}{\SU_{nom}} \geq \frac{1}{1+\theta}
	\end{eqnarray}
	where the lower bound is tight for any $v_i\leq \frac{\theta}{1+\theta}$ 
	with $i\geq 2$.

	\item \textit{\textbf{B3:} (NE Net Utility over Maximum)}
	Define a variable $\tilde{v}^{(n)}$ that is the unique, positive and real solution to the equation 
	{\setlength\abovedisplayskip{1pt plus 3pt minus 2pt} 
		\setlength\belowdisplayskip{1pt plus 3pt minus 2pt} 
		\begin{eqnarray}
		2(n{-}1)v^3 + (3-2n+\theta^{-2})v^2 - 1=0 
		\label{eq:theorem_lowerbounds_sv1}
		\end{eqnarray}}
	for any  $n\in\{2,\cdots,N\}$.
	The lower bound of the total net utility is given by \eqref{eq:theorem_lowerbounds_sv2}.
		\begin{eqnarray}
		\frac{\SV_{mal}}{\SV_{max}} &\geq& \min\{ 1- \frac{(N{-}1)((N{+}1) -\sqrt{N^2-1})}{1+\frac{1}{2}\big(\sqrt{N^2{-}1}+(N{-}1)\big)}, \nonumber\\
		&&\frac{(1-n\theta)(1+(n-1)\tilde{v}^{(n)})}{1+n\theta} \nonumber\\
		&&+ (1{+}\frac{n-1}{\tilde{v}^{(n)}})
		(\frac{\theta(1+(n-1)\tilde{v}^{(n)})}{1+n\theta})^2\}, \nonumber\\
		&& \qquad\qquad\qquad\qquad\qquad n=1,{\cdots},N.
		\label{eq:theorem_lowerbounds_sv2}
		\end{eqnarray}
	When $N$ and $\theta$ are large enough, the lower bound is approximated by 
	\begin{eqnarray}
	\frac{\SV_{mal}}{\SV_{max}} \geq \min \{\frac{1}{N}, \frac{N}{(1+N\theta)^2}\}.
	\end{eqnarray}
	
	\item \textit{\textbf{B4:} (NE Net Utility With/Without Misbehaving User)}
	{\setlength\abovedisplayskip{1pt plus 3pt minus 2pt} 
		\setlength\belowdisplayskip{1pt plus 3pt minus 2pt} 
		\begin{eqnarray}
		\frac{\SV_{mal}}{\SV_{nom}} \geq \frac{1}{(1+\theta)^2}
		\label{eq:theorem_SVmalSVnom_bound}
		\end{eqnarray}}
	where the equality holds for $v_i\leq \frac{\theta}{1+\theta}$ with $i\geq 2$.

    \item \textit{\textbf{B5:} (Operator's Revenue over Maximum)} 
    \begin{eqnarray}
    \frac{\SW_{mal}}{\SW_{max}}  \geq \frac{\theta N}{(1+\theta)(N-1)}
    \end{eqnarray}
    where $\SW_{max}$ is the maximum revenue at any NE in the absence of $\SA_0$, and the equality 
    holds for $v_i\leq \frac{\theta}{1+\theta}$ with $i\geq 2$.
     
    \item \textit{\textbf{B6:} (Operator's Revenue With/Without Misbehaving User)} 
    \begin{eqnarray}
    \frac{\SW_{mal}}{\SW_{nom}} \geq \max\{1, \frac{\theta N^2}{(1+\theta N)(N-1)}\}
    \end{eqnarray}
	where the lower bound is achieved either without the participation of misbehaving user or 
	with $v_i=1$ for all benign users. 
\end{itemize}
\label{theorem:all_lowerbounds_linear}
\end{theorem}
\noindent\textbf{Proof:} Please refer to the appendix. \done

We summarize our main observations as below. 

(1) \emph{Trend of utility dissipation.} The lower bounds of 
\textbf{B1}$\sim$\textbf{B4} decrease as the number of benign users or 
the willingness factor of the misbehaving user increases. This implies that a more furious 
competition makes the lower bounds worse off. 

(2) \emph{Role of misbehaving user in utility dissipation.} 
The utility and the net utility of benign users dissipate for two reasons: the competition 
among benign users, and that between the misbehaving user and benign users. 
When $\theta$ is small, the lower bounds \textbf{B1} and \textbf{B3} 
are governed by the number of benign users. 
In a highly competitive scenario (i.e. a large $N$), 
the lower bound of utility reduces to $\frac{3}{4}$ of the optimality and that of net utility 
is inversely proportional to $N$. 
When $\theta$ is above a certain threshold (as a function of $N$), the lower bound 
\textbf{B1} is solely determined by $\theta$ and this bound is tight 
in the presence of only $\SA_{0}$ and $\SA_{1}$ at the NE. 
The lower bound of \textbf{B3} contains both $N$ and $\theta$ because 
the worst net utility happens when as many as possible players participate 
at the NE and the willingness factor is large meanwhile. 
For the ratios \textbf{B2} and \textbf{B4}, the worst cases take place 
when $\SA_{0}$ and $\SA_{1}$ participate at the NE while other benign users 
are excluded. 
From the perspective of misbehaving user, the malicious behavior taken by 
him is effective to neutralize the utilities of benign users 
when the number of benign users is small or the valuations of most of them are small enough. The incentive of performing attacks in a highly competitive 
scenario is thus weak. 
       
(3) \emph{``Cliff effect'' of utility dissipation.}
A crucial question is how the utility of benign users is influenced by 
the bid of the misbehaving user and the network parameters. 
Instead of examining each NE case by case, we analyze the sensitivity of 
bounds to $N$ and $\theta$.
The bounds of \textbf{B1} and \textbf{B3} are sensitive to the increase of 
$N$ when $\theta$ is small. Especially, a gently increase of $N$ 
results in remarkable reduction of the lower bounds. 
When $\theta$ is greater than a certain threshold in each case, the lower bounds 
are sensitive to the change of $\theta$, especially in the vicinity of the 
threshold value. 
The lower bounds of \textbf{B2} and \textbf{B4} are decreasing and strictly 
convex with regard to $\theta$. Hence, even a very small $\theta$ is effective in 
reducing the utility and the net utility of bening users in the worst case. 

(4) \emph{More revenues to the network operator.}
The lower bound of \textbf{B6} manifests that 
the network operator can always harvest more revenues 
when the misbehaving user bids to acquire a positive amount of resources. 
This can partly interpret why many network operators simply 
ignore the misbehaving actions. The lower bound \textbf{B6} is a 
decreasing function of $N$ and an increasing function of $\theta$. 
Therefore, allowing the misbehaving user may 
generate a much better revenue to the network operator when the population of 
benign users $N$ is small or the willingness factor $\theta$ is large. 
Note that $\SW_{max}$ in \textbf{B5} is the maximum revenue obtained by the network operator 
at any NEs when the misbehaving user is excluded. As $\theta$ is small, 
the lower bound almost increases linearly to $\theta$, which might be sufficiently ``attractive'' 
to the network operator.

\noindent\textit{Remark 2:} \emph{(Challenges of Proofs)} 
Intuitively, all the theoretic bounds can be proved by simply dividing 
two metrics for each performance metric. 
However, the number of benign users 
participating at the NE with the misbehaving user might be different 
from that at the new NE without the misbehaving user. Thus, a direct division 
of two metrics at different NEs does not work. 
The participation of the misbehaving user at the NE needs to be scrutinized and analyzed 
separately,  
which greatly complicates the search of performance limits. Given the willingness factor and 
the same set of valuations in the linear 
utility function scenario, we may obtain $n_{nom}$ and $n_{mal}$ benign users participating in the 
different NEs. Two cases, $n_{nom}=n_{mal}$ and $n_{nom}\neq n_{mal}$, are considered separately; the conditions of the valuations reaching each case are investigated in the analyses. 

The upper bounds of the measures are also crucial to quantify the efficiency of 
price-anticipating Kelly mechanism. 
Firstly, the gap between an upper bound and a lower bound manifests the maximum efficiency loss 
caused by the competition among benign users and caused by the presence of the misbehaving user. 
If the willingness factor of the misbehaving user is situated in a range causing salient utility gaps, his malicious actions 
are deemed fairly effective. He is even inclined to setting the willingness factor in this region if it is adjustable. 
For a given distribution of valuations, knowing the upper bounds 
will be helpful to gauge to what extent the utility and the net utility of benign users are sacrificed, and 
the revenue of the network operator is influenced due to the competition among the benign and misbehaving users. 
Secondly, an upper bound of the revenue of network operator reflects the maximum achievable gain if 
he ``closes his eyes'' on misbehaving actions. 
Note that linear utility is a benchmark utility model in contest theory, and corresponds to the metrics such as users' 
satisfaction of using the allocated bandwidth in communication networks, 
or the users' valuation toward the received attention in online visibility competition.
The following theorem states the upper bounds of performance measures with linear utility functions.

\begin{theorem} \textit{(Upper Bounds)} 
	For linear utility functions of benign users, the upper bounds of \textbf{B1}, \textbf{B2}, \textbf{B3},  \textbf{B5} and 
	\textbf{B6} are as follows.
	\begin{itemize}
		
		\item \textit{\textbf{B1:} (NE Utility over Maximum)}
		\begin{eqnarray}
		\frac{\SU_{mal}}{\SU_{max}} \leq \left\{\begin{matrix}
		\;1 \; &&\textrm{ if } \;\; \theta \leq \frac{N-1}{N} \\
		\; \frac{N}{\theta N{+}1} \;  &&\textrm{ otherwise }
		\end{matrix}\right.
		\label{eq:theorem_no4}
		\end{eqnarray}
		where the upper bound is tight for $v_i=1$ $(i\geq 1)$. 
		
		\item \textit{\textbf{B2:} (Approximate Upper Bound of NE Utility With/Without Misbehaving User)} 
		Given linear utility functions of the benign users, the upper bound of $\frac{\SU_{mal}}{\SU_{nom}}$ is approximated by 
		{\setlength\abovedisplayskip{1pt plus 3pt minus 2pt} 
			\setlength\belowdisplayskip{1pt plus 3pt minus 2pt} 
			\begin{eqnarray}
			\!\!\!\!\!\!\frac{\SU_{mal}}{\SU_{nom}} {\leq} \left\{\begin{matrix}
			1  \quad\quad \textrm{if } \theta \leq \frac{N{-}1}{N}\\
			\max\{\frac{N}{1{+}N\theta}, \frac{(1+\theta)^{-1}}{(1-(N-1)(\sqrt{N}-\sqrt{N-1})^2)}\} \;\; \textrm{otherwise }
			\end{matrix}\right.
			\label{eq:theorem_SU_malvsnom_upper}
			\end{eqnarray}}
		where it is tight when $v_i$ equals to 1 for all $i$. 
				
		\item \textit{\textbf{B3:} (NE Net Utility Over Maximum)}
		{\setlength\abovedisplayskip{1pt plus 3pt minus 2pt} 
			\setlength\belowdisplayskip{1pt plus 3pt minus 2pt} 
			\begin{eqnarray}
			\frac{\SV_{mal}}{\SV_{max}} \leq \left\{\begin{matrix}
			\frac{1}{(1+\theta)^2} && \textrm{ if } \theta\leq\sqrt{2}{-}1 \\
			\frac{1}{2} && \textrm{ if } \sqrt{2}{-}1 < \theta\leq\frac{1}{2} \\
			\frac{2}{(1+2\theta)^2}, && \textrm{ if } \; \frac{1}{2}<\theta\leq\frac{\sqrt{2}}{2}\\
			\frac{1}{(1+\theta)^2}, && \textrm{ if } \; \theta >\frac{\sqrt{2}}{2}
			\end{matrix}\right..
			\label{eq:theorem_no5}
			\end{eqnarray}}
		
    \item \textit{\textbf{B5:} (Operator's Revenue over Optimality)} 
    \begin{eqnarray}
    \frac{\SW_{mal}}{\SW_{max}} \leq \max\{1, \frac{\theta N^2}{(1+\theta N)(N-1)}\}
    \end{eqnarray}
    where the upper bound is tight for $v_i=1$ $(i\geq 1)$. 
    \item \textit{\textbf{B6:} (Operator's Revenue With/Without Misbehaving User)} 
    \begin{eqnarray}
    \frac{\SW_{mal}}{\SW_{nom}} \propto \infty
    \end{eqnarray}
    where the unbounded ratio takes place asymptotically as $v_2$ approaches 0. 
    \end{itemize}
\label{theorem:all_upperbounds_linear}
\end{theorem}
\noindent\textbf{Proof:} Please refer to the appendix. \done

Our major observations on the upper bounds are summarized below. 

\emph{(1)  Trend of utility dissipation and cliff effect.} 
The upper bounds of \textbf{B1} and \textbf{B2} reach 1
when the willingness factor is small. As it further increases beyond a certain threshold, 
even a small increment can yield a remarkable reduction on the upper bounds. 
When $N$ is sufficiently large, the asymptotic upper bound of  \textbf{B1} is inversely proportional to 
the willingness factor $\theta$, 
and that of \textbf{B2} is approximated as $\max\{\frac{1}{\theta}, \frac{4}{3(1+\theta)}\}$.
Similarly, the upper bound of \textbf{B3} is very sensitive to the willingness factor. 
The piece-wise curves indicate the participation of users in four scenarios: \{$\SA_0$, $\SA_1$\}, \{$\SA_1$, $\SA_2$\}, 
\{$\SA_0, \SA_{1}, \SA_{2}$\}, and \{$\SA_0, \SA_1$\} when the upper bound of \textbf{B3} is obtained.

\emph{(2) Surprising revenue gain to the network operator.} The misbehaving user is inclined to bring 
surprisingly high revenue gains. The ratio \textbf{B5} reflects the maximum achievable revenue at any NE. 
One can observe that the presence of the misbehaving user is able to generate a revenue higher than 
the maximum revenue without him. The ratio \textbf{B6} compares the revenues with and without the misbehaving user at the 
NEs. The presence of the misbehaving user brings an unbounded gain of the revenue at the extreme situation 
that all the benign users except the first one own very low valuations. This implies that the misbehaving user 
is especially beneficial to the network operator when the competition among benign users is gentle. 

\subsection{General Utility Functions and Multiple Misbehaving Users}

The linear utility function is of the simplest form. An important question is 
whether the theoretic bounds derived above are applicable beyond linear utility functions. Following the approach in \cite{johari}, we show that 
the lower bounds of linear utility serve as those of more general utilities 
satisfying \emph{Assumption \textbf{P}}.

\begin{theorem} \textit{(Condition for Lower Bounds)} 
	For arbitrary utility functions of the benign users satisfying \emph{Assumption \textbf{P}}, 
	the performance metrics, $\frac{\SV_{mal}}{\SV_{max}}$, $\frac{\SU_{mal}}{\SU_{nom}}$ and
	$\frac{\SV_{mal}}{\SV_{nom}}$, are lower bounded by the cases with linear utility functions.
	\label{theorem:no3_lowerboundscondition}
\end{theorem}
\noindent\textbf{Proof:} Please refer to the appendix. \done



A natural extension of our study is to consider multiple misbehaving users. 
Denote by $\SK=\{1, \cdots, K\}$ the set of misbehaving users. Let $\theta_k$ be the willingness to pay of the 
$k^{th}$ misbehaving user, and let $x_{0k}$ be his bid paid to the network operator. 
The NE conditions for the misbehaving users are given by 
\begin{eqnarray}
\theta_k \sum\nolimits_{i{=}1}^{N} U_{\SA_i}'(d_i^*) \frac{x_{i}^*}{(z^*)^2} - 1\left\{\begin{matrix}
=0   &\textrm{ if }  x_{0k}^* > 0 \\
\leq 0 &\textrm{ if }   x_{0k}^* = 0 
\end{matrix}\right., \quad \forall k
\label{eq:onetarget_multi_misbehaving}
\end{eqnarray}
according to the principle of KKT conditions 
where $z^* = \sum\nolimits_{j=1}^{K}x_{0j}^* + \sum\nolimits_{j=1}^{N}x_{j}^*$. 
Consider two misbehaving users $k_1$ and $k_2$ with $\theta_{k_1}> \theta_{k_2}$. 
We suppose that the equalities hold for both $k_1$ and $k_2$ in Eq.\eqref{eq:onetarget_multi_misbehaving} when they pay positive bids. 
Subtracting one from the other on both sides, we obtain 
$(\theta_{k_1}-\theta_{k_2})\big(\sum\nolimits_{i{=}1}^{N} U_{\SA_i}'(d_i^*) \frac{x_{i}^*}{(z^*)^2}\big) = 0$. Given $\theta_{k_1}> \theta_{k_2}$ 
and $(\sum\nolimits_{i{=}1}^{N} U_{\SA_i}'(d_i^*) \frac{x_{i}^*}{(z^*)^2}\big) >0$, the above equality does not hold, which contradicts the assumption. 
Hence, only the misbehaving user with the larger willingness factor 
may participate at the NE. For a set of misbehaving users with heterogeneous 
willingness factors, the one with the largest willingness factor 
can pay a positive bid, while the others do not participate in the competition.

When multiple misbehaving users possess the identical willingness factor (or when the misbehaving user splits himself into multiple users), the uniqueness of the NE cannot be guaranteed. 
The NE conditions for the misbehaving users are given by 
\begin{eqnarray}
\theta \sum\nolimits_{i{=}1}^{N} U_{\SA_i}'(d_i^*) \frac{x_{i}^*}{(z^*)^2} - 1\left\{\begin{matrix}
=0   &\textrm{ if }  x_{0k}^* > 0 \\
\leq 0 &\textrm{ if }   x_{0k}^* = 0 
\end{matrix}\right., 
\label{eq:splitting}
\end{eqnarray}	
for every misbehaving user.  The KKT conditions have a unique solution to $z^*$, 
yet the combinations of $\{x_{0k}\}_{k=1}^K$ 
resulting in the unique $z^*$ are infinite. Therefore, the existence of multiple misbehaving users 
affects the Kelly mechanism in the same way as a single misbehaving user.

\subsection{Impact of Incomplete Information}

Nash equilibrium is obtained on the basis of complete information, i.e. each player knows 
the utility functions and the strategy profiles of all the players. 
Because our purpose is to uncover the efficiency loss at the NE, we assume the 
availability of complete information. In practice, a benign user is only aware of his own 
utility function and is unwilling to share this information to all the opponents. 
In this situation, the game is played many rounds so that the complete information can be 
learned implicitly by each user. 
Here, one round corresponds to a time slot (e.g. one second) 
of sampling flow throughput in DDoS attack, or  
a time slot of visibility competition (e.g. one day) in online media. 
Both the flow transmission and the online promotion take a number of rounds 
so that the equilibrium could be reached. 
Each benign user $\SA_i$ knows his utility function, his bid $x_i$ and his allocated resource $d_i=\frac{x_i}{z}$  
at time $t$. Given this knowledge, he is trying to maximize his net utility in the next round. 
The best response of $\SA_i$ is to choose $x_i(t+1)$ that gives rise to the largest $U_{\SA_i}(d_i)$ for $i\geq 1$, 
\begin{eqnarray}
x_i(t+1)  =  \max\{x_i(t)|_{U'_{\SA_i}(d_i(t))\cdot \frac{z_{-i}(t)}{(z(t))^2} = 1}, \; 0\}. 
\end{eqnarray} 
The goal of the misbehaving user is to mitigate the utilities of benign users so that 
these information is known priori or can be inferred 
so that he is capable of launching misbehaving actions. 
The best response of $\SA_0$ is expressed as 
\begin{eqnarray}
x_0(t+1)  =  \max\{x_0(t)|_{\sum_{i=1}^{N}U'_{\SA_i}(d_i(t))\cdot \frac{x_i(t)}{(z(t))^2} = \frac{1}{\theta}}, \; 0\}, 
\end{eqnarray} 
which means that $U_{\SA_i}(d_i)$ $(\forall i\geq 1)$ can be observed or inferred by $\SA_0$. 
This is feasible in visibility competition and crowdsensing applications because 
the bids of a user in the form of promotion messages and contributed efforts are directly observable 
by the misbehaving user. 
In other applications such as DDoS attack, $\SA_i$'s bid, $x_i$, is not easily acquired, while the 
misbehaving user can still infer the global utility through perceiving the 
sensibility of the system by producing attacks. 

We hereby provide a concrete example of best response functions where 
$U_{\SA_i}(d_i)$ is a logarithmic function taking the form $U_{\SA_i}(d_i) = \log(1 + d_i)$. 
The above best response function is given by 
\begin{eqnarray}
x_i(t+1)  =  \max\{\frac{-3z_{-i}(t) + \sqrt{8z_{-i}(t) + (z_{-i}(t))^2}}{4}, \; 0\}. 
\end{eqnarray} 
Here, $z_{-i}(t)$ is the total bid of other users, which is observed directly or can be deduced from the 
acquired resources by $\SA_i$. 
Given the fixed $x_0$, each benign user can update his strategy in each step and may converge to the Nash 
equilibrium distributedly. Therefore, the incomplete information does not prevent the search of NE by the benign users 
for a fixed $x_0$. However, when $x_0$ is determined strategically by the misbehaving user, 
he needs to know the utility functions of the benign agents since his goal is to minimize their total utility. 
Hence, the best response of the misbehaving user is 
\begin{eqnarray}
x_0(t+1)  =  \max\{ \sqrt{\theta \sum\nolimits_{i=1}^{N}\frac{x_i(t)}{1+d_i(t)}} - z_{-0}, \; 0\}. 
\end{eqnarray}

\section{Numerical Results}
\label{sec:simu}

In this section, we validate the proposed bounds through numerical simulations and reveal the 
important insights on the efficiency of Kelly mechanism in adversity.

\begin{figure}[!htb]
		\centering
		\includegraphics[width=2.8in]{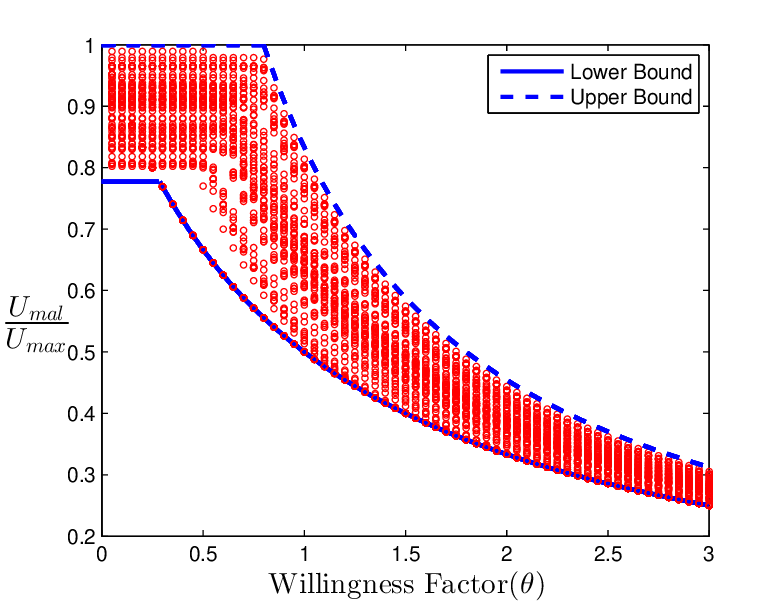}
		\caption{Bounds of $\frac{\SU_{mal}}{\SU_{max}}$ with linear utility functions}
		\label{fig:linearbd_UmalUmax}
\end{figure}

\begin{figure}[!htb]
		\centering
		\includegraphics[width=2.8in]{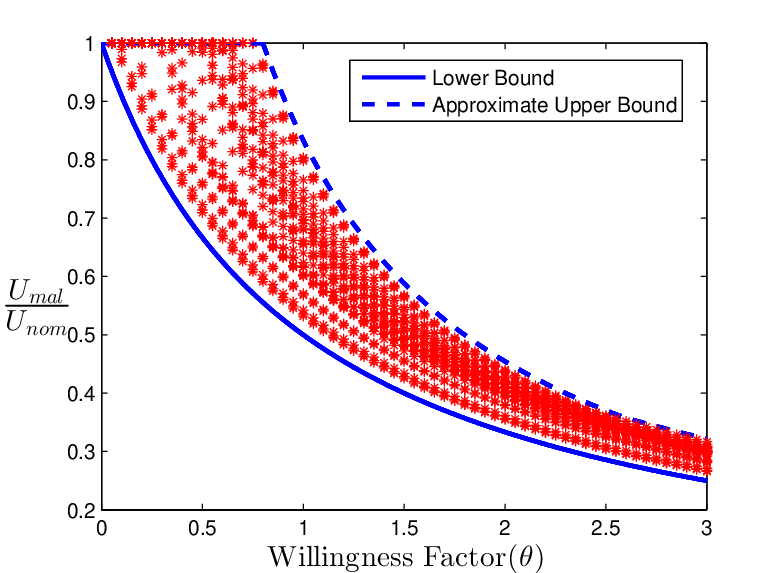}
		\caption{Bounds of $\frac{\SU_{mal}}{\SU_{nom}}$ for linear utility functions}
		\label{fig:linearbd_UmalUnom}
\end{figure}

\begin{figure}[!htb]
		\centering
		\includegraphics[width=2.8in]{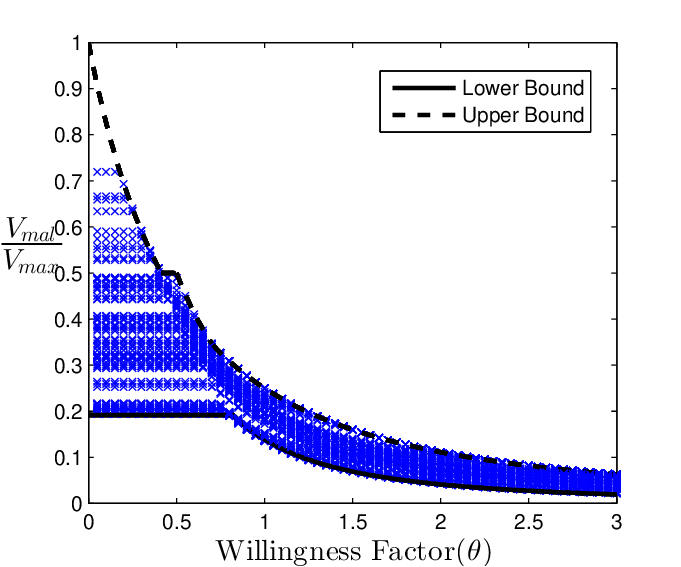}
		\caption{Bounds of $\frac{\SV_{mal}}{\SV_{max}}$ for linear utility functions}
		\label{fig:linearbd_VmalVmax}
\end{figure}

\begin{figure}[!htb]
		\centering
		\includegraphics[width=2.8in]{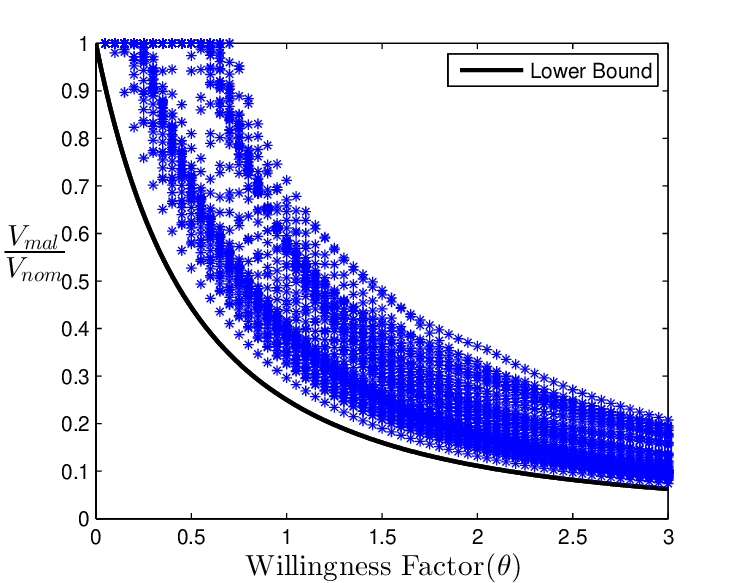}
		\caption{Lower Bound of $\frac{\SV_{mal}}{\SV_{nom}}$ for linear utility functions}
		\label{fig:linearbd_VmalVnom}
\end{figure}

\begin{figure}[!htb]
		\centering
		\includegraphics[width=3in]{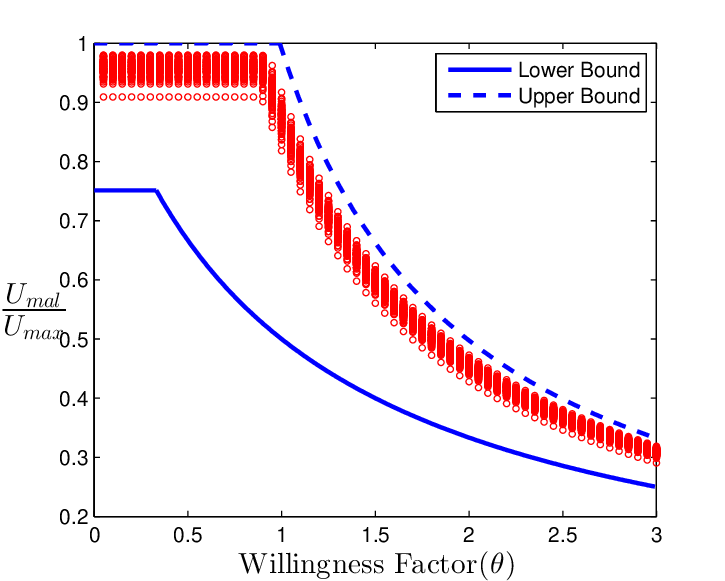}
		\caption{Bounds of $\frac{\SU_{mal}}{\SU_{max}}$ for linear utility functions ($N=100$)}
		\label{fig:Logbd_WmalWmaxLog_N100}
\end{figure}

\noindent\textbf{Linear Utility Functions:} 
We hereby compute the performance measures for a large number of random tests. 
In Fig.\ref{fig:linearbd_UmalUmax}$\sim$\ref{fig:Logbd_WmalWmaxLog_N100}, we consider the timeline 
competition among five benign users ($N=5$) and one misbehaving user. Here, x-coordinate denotes 
the willingness factor, and y-coordinate denotes different ratios respectively. 
Each marked point in these figures with a marker represents a corresponding ratio with randomly 
generated valuations from uniform distribution in [0, 1] 
(some points are not plotted simply for reducing the size of image files). 
Fig.\ref{fig:linearbd_UmalUmax} illustrates the upper and the lower bounds of $\frac{\SU_{mal}}{\SU_{max}}$.
The misbehaving user does not participate in the worst case of $\SU_{mal}$ when 
$\theta$ is less than 0.287. As $\theta$ further increases, the lower bound descends rapidly. 
Similarly, the misbehaving user does not participate in the best case of $\SU_{mal}$ 
when $\theta$ is less than 0.8. A slight increase of $\theta$ beyond 0.8 leads to a rapid decrease 
of the upper bound. 
Thus, for the small willingness factor, we can deem that the efficiency loss in the total utility 
is mainly caused by the competition among the benign users. The bounds of utility loss 
are very sensitive to the change of $\theta$ in certain ranges. 
Fig.\ref{fig:linearbd_UmalUnom} compares the utilities at two NEs for the games 
$\mathbf{MAL}$ and $\mathbf{NOM}$. The lower bound is shown to decrease rapidly even 
for very small $\theta$. 
Fig.\ref{fig:linearbd_VmalVmax} illustrates the upper and the lower bounds of $\frac{\SV_{mal}}{\SV_{max}}$. 
The upper bound of the net utility is shown to be very sensitive to $\theta$, especially when 
$\theta$ is small. This implies that the best net utility is not robust against the misbehaving 
actions. The lower bound is fixed with small $\theta$, but decreases when $\theta$ is 
above a certain threshold. This manifests that the loss of net utility is caused by the 
competition among the benign users for small $\theta$ and by the misbehaving user 
for large $\theta$. The lower bound of $\frac{\SV_{mal}}{\SV_{nom}}$ is shown in 
Fig.\ref{fig:linearbd_VmalVnom}, which owns a similar property as the lower bound of 
$\frac{\SU_{mal}}{\SU_{nom}}$ in Fig.\ref{fig:linearbd_UmalUnom}. 
As $\theta$ increases from 0, the lower bound decreases quickly.

\begin{figure}[!htb]
		\centering
		\includegraphics[width=2.8in]{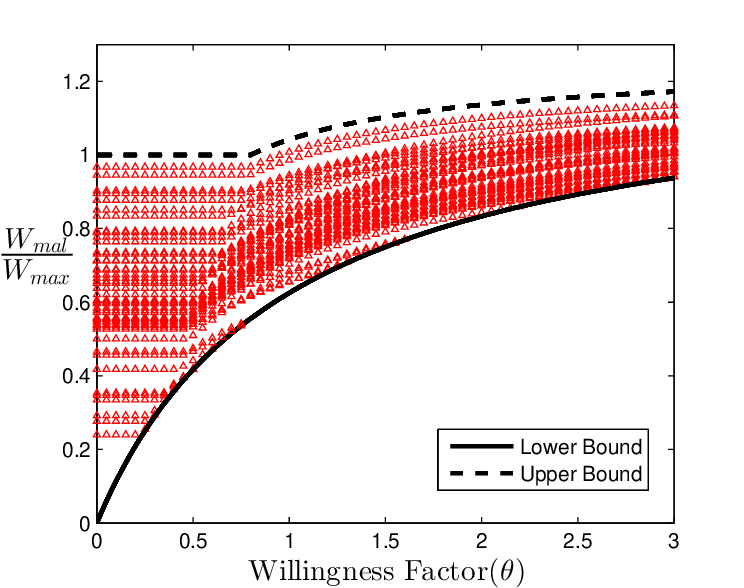}
		\caption{Bounds of $\frac{\SW_{mal}}{\SW_{max}}$ for linear utility functions}
		\label{fig:linearbd_WmalWmax}
\end{figure}

\begin{figure}[!htb]
		\centering
		\includegraphics[width=2.8in]{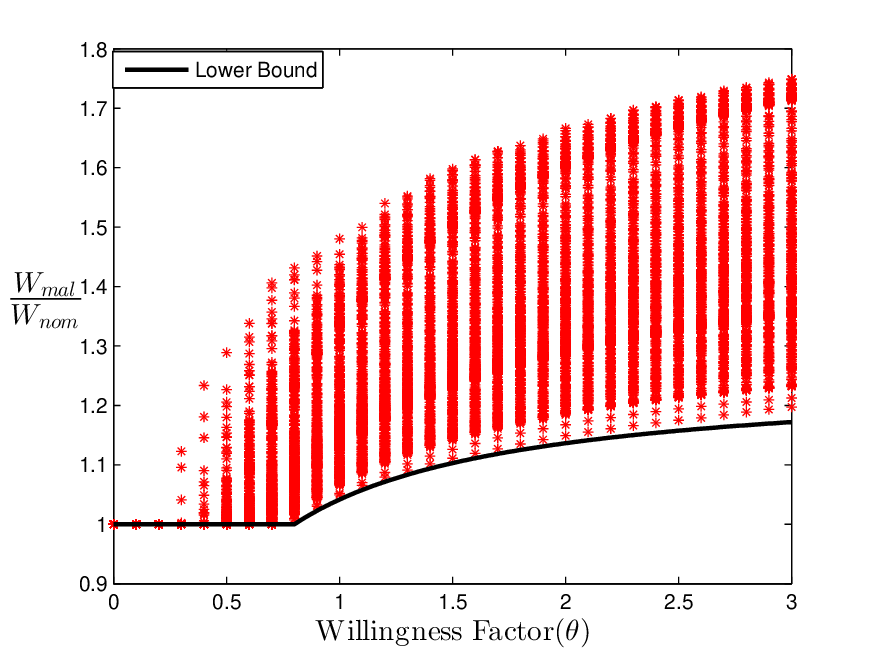}
		\caption{Bounds of $\frac{\SW_{mal}}{\SW_{nom}}$ for linear utility functions}
		\label{fig:linearbd_WmalWnom}
\end{figure}

Fig.\ref{fig:linearbd_WmalWmax} plots the upper and the lower bounds of $\frac{\SW_{mal}}{\SW_{max}}$. 
When $\theta$ increases from 0, the lower bound increases accordingly. When $\theta$ is greater than 
$0.8$, the misbehaving user is bound to participate. It is interesting to see 
that the participation of the misbehaving user brings more revenues to the network operator. 
When $\theta$ is 3, the upper bound of the game $\mathbf{MAL}$ is nearly 20\% higher than 
the best revenue in the game $\mathbf{MAX}$. 
Fig.\ref{fig:linearbd_WmalWnom} compares the revenue of the network operator at the NEs 
$\mathbf{MAL}$ and $\mathbf{MAX}$. The revenue of the network operator in $\mathbf{MAL}$ is invariably 
no less than that in $\mathbf{MAX}$. Especially, the maximum improvement in our simulation is 
more than 70\% when $\theta$ is 3. In light of the benefits brought to the network operator, he may  
renounce the regulation of the misbehaving user in targeted advertising. 
Fig.\ref{fig:Logbd_WmalWmaxLog_N100} plots the competition with $N=100$ benign users. 
The valuations of benign users are identically independent distributed in [0, 1]. 
Hence, some of them are close to 1 when $N$ is large so that the ratio $\frac{\SU_{mal}}{\SU_{max}}$
is kept at a high level. 
Compared with Fig.\ref{fig:linearbd_UmalUmax}, 
the results in Fig.\ref{fig:Logbd_WmalWmaxLog_N100} are more convergent and closer to the upper bound. 
Note that the lower bound at $N=100$ is reached when all the valuations in the set $\{v_i\}_{i=2}^N$ 
are uniformly chosen as 0.499. 
Since each $v_i$ ($i\geq 2$) is randomly selected in the range $[0, 1]$, trying a limited number of tests can hardly 
reach a set of $\{v_i\}_{i=2}^N$ that yield the worst ratio \textbf{B1}. Actually, a lot of benign users possess relatively large valuations for a large $N$, causing the corresponding samples of the ratio \textbf{B1} much higher 
than the lower bound. This also reflects the robustness of price-anticipating Kelly mechanism in terms of the utility of benign users at the large population scenario. 




\begin{figure}[!htb]
		\centering
		\includegraphics[width=2.8in]{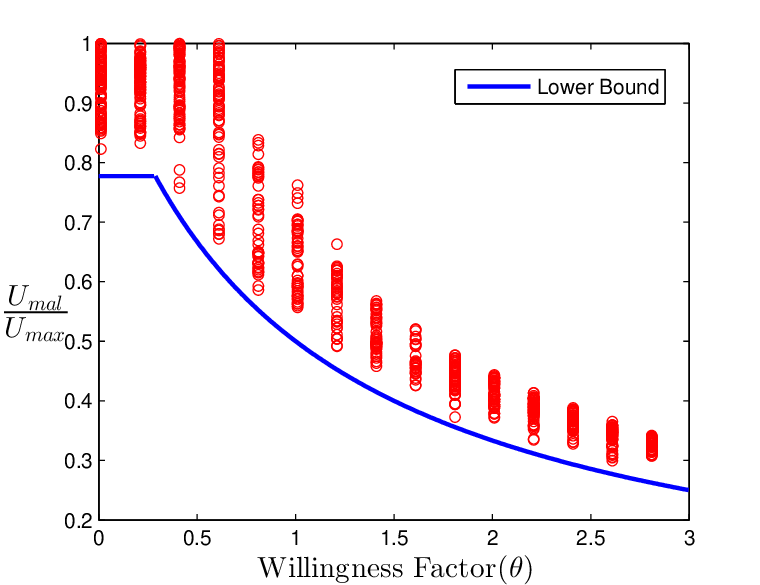}
		\caption{Bounds of $\frac{\SU_{mal}}{\SU_{max}}$ for logarithmic utility functions}
		\label{fig:Logbd_VmalVnom}
\end{figure}

\begin{figure}[!htb]
		\centering
		\includegraphics[width=3in]{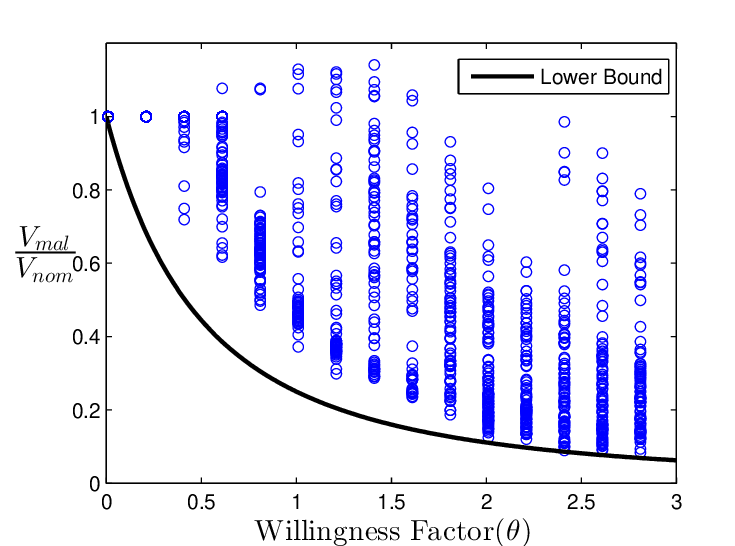}
		\caption{Lower Bound of $\frac{\SV_{mal}}{\SV_{nom}}$ for logarithmic utility functions}
		\label{fig:Logbd_VmalVnomLog}
\end{figure}

\noindent\textbf{Logarithmic Utility Functions:}
We evaluate the lower bounds for logarithmic utility functions in the form 
$U_{\SA_i}(d_i) = a_i\log(1+b_id_i)$ where $a_i$ and $b_i$ are random positive parameters 
in the range [0, 1]. 
The number of benign users is set to five. 
Fig.\ref{fig:Logbd_VmalVnom}$\sim$\ref{fig:Logbd_VmalVnomLog} demonstrate the 
correctness of the lower bounds for general utility functions. 
One can observe that in each set of experiments, the lower bound is 
sensitive to the change of the willingness factor in a certain range. 
Note that the upper bounds of the scenario with linear utility functions do not apply to 
the one with logarithmic utility functions.

\begin{figure}[!htb]
		\centering
		\includegraphics[width=2.8in]{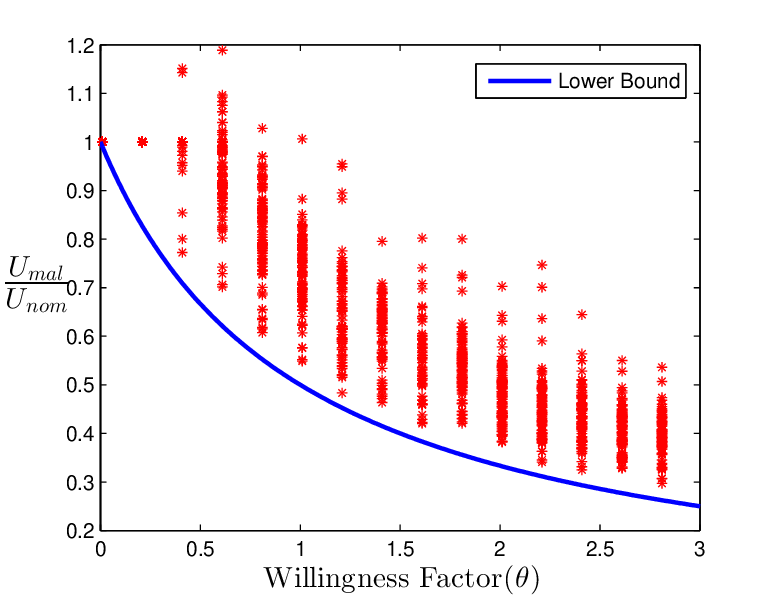}
		\caption{Bounds of $\frac{\SU_{mal}}{\SU_{nom}}$ for logarithmic utility functions}
		\label{fig:Logbd_WmalWmaxLog}
\end{figure}

\begin{figure}[!htb]
		\centering
		\includegraphics[width=2.8in]{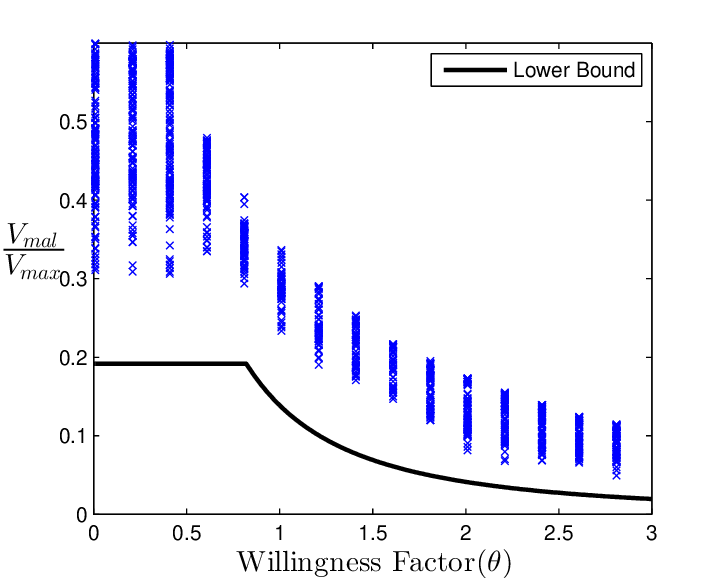}
		\caption{Bounds of $\frac{\SV_{mal}}{\SV_{max}}$ for logarithmic utility functions}
		\label{fig:Logbd_WmalWnomLog}
\end{figure}

\section{Countermeasure}
\label{sec:implications}

In this section, we evaluate the impact of differential pricing on the efficiency loss of 
Kelly mechanism in which the network operator charges different prices of per unit of 
bid from the players.

\subsection{Differential Pricing}

The bid of a user has multifarious physical interpretations in different applications. 
For instance, the bid usually refers to the payment to network operator in the bandwidth allocation game \cite{johari}; 
it also refers to the sending rate of advertising messages in online visibility competition \cite{Altman1}. 
The standard Kelly mechanism does not differentiate who contributes a bid so that the ``per unit price'' of all bids is deemed as 1 uniformly. In practice, the network operator can steer the bidding of players 
by charging them different prices where this generalization of Kelly mechanism is named 
``price differentiation'' \cite{yang2013price}. 
We choose the online visibility competition as a \emph{use case} to avoid confusion, 
where the benign agents aim to gain visibility through online advertising and the misbehaving agent, usually 
a market incumbent, tries to mitigate the visibility of benign agents. 
The fraction of visibility that a user obtains is proportional to his message sending rate, but inversely 
proportional to the aggregate message sending rate, and the benign users have different 
valuations on each unit of visibility \cite{Altman1}. We suppose that $\SA_i$ is charged 
at a price $c_i (c_i\geq 1)$ for generating advertising messages by a normalized rate to gain eyeballs, 
and the fraction of harvested visibility is $d_i$ in 
Eq.\eqref{eq:d_i}. 
Then, the cost incurred by $\SA_i$ is thus $c_i x_i$ if the message sending rate is $x_i$. 
The price differentiation scheme is able to steer the resource allocation toward a better network utility. 

Our analysis on the robustness of Kelly mechanism manifests that the misbehaving user has the potential to 
significantly reduce network utility. To counteract the adversarial actions, the network operator 
can wield the ``weapon'' of differential pricing against the misbehaving user. 
An interesting question arises: \emph{how will the differential pricing influence the utility and the net utility of benign users, the 
revenue of the network operator, and their lower bounds?}
We hereby consider three scenarios separately: the uniform pricing, the differential pricing on the misbehaving user, and the 
improper differential pricing on the benign user \footnote{We claim the differential pricing on a benign user as an \emph{improper} 
pricing because it is unfair to penalize him in our context.}. 

\subsection{Impact of Differential Pricing on Kelly Mechanism}

\subsubsection{Uniform Pricing}
All the users are charged the same price $c$ of generating advertising messages at a unit rate. 
When using the strategy $x_i$, $\SA_i$ pays a fee $cx_i$ to the network operator. 
For the linear utility functions of benign users, the NE retains the same forms as 
Eq.\eqref{eq:linearutility_no1} and \eqref{eq:linearutility_no2} except that $v_i$ is 
substituted by $\frac{v_i}{c}$. 
When $n$ benign users participate at the NE, the threshold of willingness factor remains unchanged 
so that adjusting the price $c$ does not alter the participation of the misbehaving user. 
The strategy of $\SA_i$ at the new NE shrinks by $1/c$ times for $i\in [0, N]$, and the 
payment of a user to the network operator remains unchanged (easily drawn from Theorem \ref{theorem:linear_ne}). 
Therefore, the fraction of resources allocated to a benign user, his utility and 
net utility do not change either. 
Meanwhile, because all the valuations are equally dimensioned, 
the bounds of \textbf{B1}$\sim$\textbf{B6} are unaltered. We reiterate the major observations.

\begin{corollary} 
	Choosing a uniform price for all the users does not change competition intrinsically. The utility of each user is left unchanged, 
	so are the lower bounds of performance measures. 
	\label{lemma:corollary_uniform_pricing}
\end{corollary}

\subsubsection{Differential Pricing on $\SA_0$}
The misbehaving user is successfully identified by the network operator and is charged at a higher price 
$c>1$ while the benign users are charged at a unit price. 
For the linear utility functions of benign users, the NE is computed from 
Eq.\eqref{eq:linearutility_no1} and \eqref{eq:linearutility_no2} with the substitution of $\theta$ by $\theta/c$. 
As $c$ increases, the actual willingness factor $\theta/c$ may fall below the threshold of participation. 

\begin{figure}[!htb]
		\centering
		\includegraphics[width=2.8in]{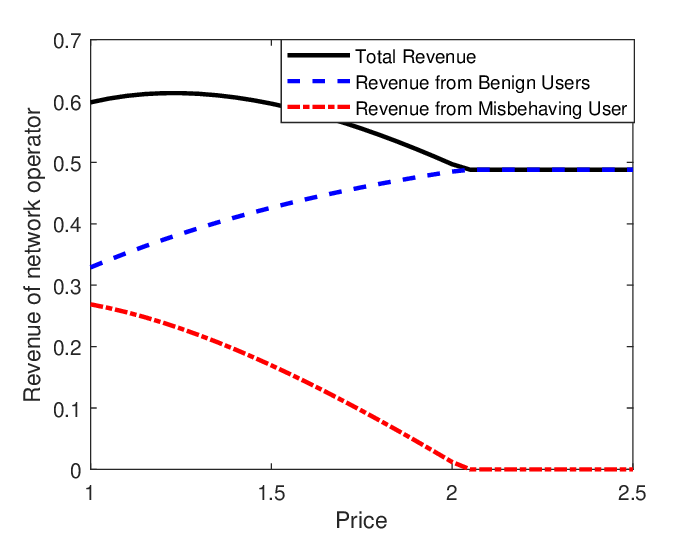}
		\caption{Total revenue and revenues from benign users as well as misbehaving user: price differentiation on $\SA_0$.}
		\label{fig:diff_pricing1}
\end{figure}

When the misbehaving user and $n$ benign users participate at the NE, 
the misbehaving user sends messages at a lower rate according to Eq.\eqref{eq:linearutility_no2}. 
The total utility of benign users is $\frac{\sum_{i=1}^nv_i}{n\theta/c +1}$, and is thus an increasing function of $c$. Similarly, the total net utility of benign users is also an increasing function of $c$ according to 
Theorem \ref{theorem:linear_ne}.
The total bid of all users, $\frac{\theta\sum_{j=1}^nv_j}{n\theta+c}$, is a decreasing function of $c$ 
where the benign users increase their bids yet the misbehaving user decreases his bid. 
It is uncertain how the revenue of the network operator is influenced by the differential pricing 
on the misbehaving user. 
We adopt two examples with five benign users and one misbehaving user to explain our observations. 
When the valuation vector is $\bv=[1, 0.66, 0.63, 0.23, 0.18]$ and the willingness factor is 
$\theta=1.2$, we plot the revenue of the network operator in Fig. \ref{fig:diff_pricing1} by 
increasing $c$ from 1 to 2.5. The revenue curve is not monotone: a slight price differentiation 
may bring more revenues to the network operator, while charging a much higher price reduces his revenue 
considerably. We next examine the impact of $c$ on the bounds of performance measures. 
Because changing $c$ is equivalent to altering $\theta$, the lower bounds retain the same forms 
in Theorem \ref{theorem:all_lowerbounds_linear}, but with the substitution of $\theta$ by $\frac{\theta}{c}$
in the bounds \textbf{B1}$\sim$\textbf{B4}.


\begin{corollary} 
	The differential pricing on the misbehaving user can effectively improve the utility and the net utility 
	of benign users, while it is usually (though not always) detrimental to the revenue of network operator when the price is high. The new lower bounds of \textbf{B1}$\sim$\textbf{B4} are equivalent to those with the willingness factor of $\frac{\theta}{c}$.
	\label{lemma:corollary_diff_pricing}
\end{corollary}

\subsubsection{Differential Pricing on a benign user.} 
\begin{figure}[!htb]
		\centering
		\includegraphics[width=2.8in]{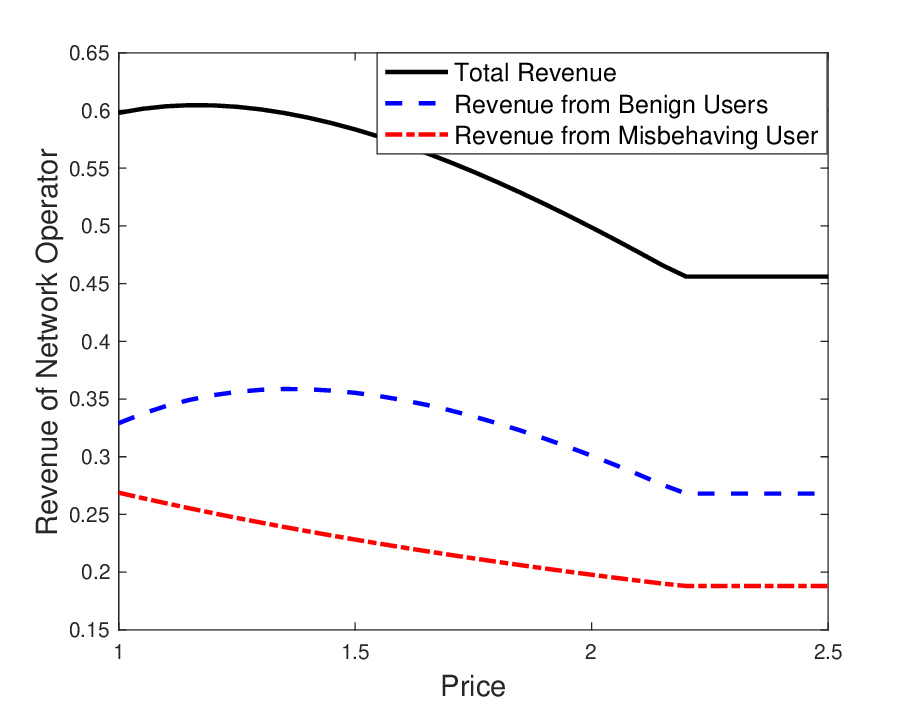}
		\caption{Total revenue and revenues from benign users as well as misbehaving user: price differentiation on $\SA_1$.}
		\label{fig:revenue_operator_wrongdiff}
\end{figure}

The differential pricing might incur a hazard that a benign user is mistaken 
as a misbehaving user. Here, we suppose $\SA_1$ is penalized by the network operator. 
We consider two scenarios with regard to whether the misbehaving user participates at the NE. 
When $n$ benign users including $\SA_1$ generate positive message rates but $\SA_0$ does not participate, 
the total utility of benign users is given by $\sum_{i=1}^{n}v_i - \frac{(n-1)(c+n-1)}{\frac{c}{v_1}+\sum_{i=2}^{n}\frac{1}{v_i}}$. Taking its derivative over $c$, one can easily show that the total utility is strictly 
decreasing with regard to $c$ until $\SA_1$ does not participate in the competition. 
When $\SA_0$ participates at the NE, the total utility of benign users is $\frac{\sum_{i=1}^{n}v_i}{(c+n-1)\theta + 1}$, a decreasing function of $c$. However, such a monotone property does not hold 
in the net utility. As $c$ increases, charging a higher price to $\SA_1$ causes him to reduce the 
message sending rate, and yields a lower total net utility of benign users. When $c$ is large than a 
certain threshold, the competition among benign users is lessened, causing an increase in the 
net utility until $\SA_1$ is completely excluded at the NE. 
For the sake of the same reason, the impact of $c$ on the revenue of network operator is not monotone, 
either. Choosing the set of valuations as $\bv=[1, 0.66, 0.63, 0.23, 0.18]$ and the willingness factor as 
$\theta=1.2$, the revenue of network operator is shown in Fig. \ref{fig:revenue_operator_wrongdiff}. 
The increase of $c$ in the very beginning encourages more benign users to increase their message 
sending rates, thus causing an improved total revenue. As $c$ further increases, penalizing $\SA_1$ 
reduces the intensity of competition, leading to the decreased revenue of network operator. 

Mischarging a benign user influences the lower bounds of Kelly mechanism considerably. We may follow 
the analyses in the proof of Theorem \ref{theorem:all_lowerbounds_linear}, though the expressions of 
lower bounds are much more complicated. 
With improper differential pricing on $\SA_1$, the lower bound of benign users' utility at the NE over the social optimum (\textbf{B1}) can be analyzed following the same approach of that with uniform pricing, and hence 
step-by-step math operations are omitted. 
When the network operator sets a price $c$ $(c>1)$ to $\SA_1$, the lower bound \textbf{B1} 
takes place at two situations: i) $\frac{\SU_{mal}}{\SU_{max}} \geq 1- \frac{(n-1)(\sqrt{n+c-1}-\sqrt{n-1})^2}{c}$ without the participation of $\SA_0$ and ii) $\frac{\SU_{mal}}{\SU_{max}} \geq \frac{1}{1+c\theta}$ with 
the participation of $\SA_0$ at the NE. 
With the increase of $c$, the lower bound of \textbf{B1} decreases at both situations. 
The implication is that the worst utility ratio $\frac{\SU_{mal}}{\SU_{max}}$ suffers from the price differentiation 
no matter whether the misbehaving user exists or not. When the misbehaving user participates at the NE, 
a smaller willingness factor can yield the same low bound of $\frac{\SU_{mal}}{\SU_{max}}$. 


\begin{corollary} 
	The differential pricing on a benign user reduces the total utility of benign users, while its impacts on the net utility of benign users and the revenue of network operator remain obscure. Meanwhile, 
the lower bound of \emph{NE Utility over Maximum} becomes smaller, thus making the misbehaving 
user more harmful to the benign users as a whole.  
	\label{lemma:corollary_diff_pricing2}
\end{corollary}

\section{Related Work}
\label{sec:related}

We describe the closely related works from three perspectives and summarize our major differences from the literature.

\emph{Timeline Competition:}  
An important feature is that popular OSNs such as Facebook, Google+ and Twitter adopt a 
timeline-based template to sort content reverse chronologically. \cite{Altman1} 
first studied the competition between different content creators on a user's timeline. 
The timeline competition is modeled as a non-cooperative game in \cite{Altman2} 
that is in the standard form of Kelly mechanism. 
They further extended 
the Kelly mechanism to the parallel competition on multiple topics in a timeline \cite{Altman3}. 
Fully dynamic and semi-dynamic models have been proposed in \cite{Altman4} and \cite{Altman5} 
respectively to characterize the competition on followers' timeline over time.

\emph{Contest Theory:} In politics and economics, there are many situations where users fight 
over property rights. 
The kernel feature of a contest is the contest success function (CSF)  
developed from the seminal work by Tullock \cite{tullock1,tullock2}. 
Rent-seeking, a specific contest, was studied in \cite{krueger} and a lobbying contest was 
studied in \cite{becker}.
The Tullock CSF is the special case of our timeline competition model when the utility 
functions are all linear and the misbehaving user does not exist. 
More recently, the studies on CSF mainly focus on the equilibrium analysis with incomplete 
information \cite{corchon,wasser} and the design of CSF mechanism \cite{franke,polishchuk}. 



\emph{Bandwidth Allocation Games:} 
F.P. Kelly proposed a market mechanism 
in which each user submits a ``bid'' per unit time to the network operator, and the network operator 
determines the share of each user at a link \cite{kelly}. In particular, the bandwidth share is proportional to his bid and 
inversely proportional to the sum of all users' bids at a single link. 
Authors in \cite{johari} established that 
the aggregate utility received by users was at least 3/4 of the maximum possible aggregate utility. 
A ``misbehaving'' player was introduced in the Kelly mechanism to study the 
lower bound of the total utility of benign users at an infinite population regime \cite{Vulimiri}.
Other important works studied the 
performance of market equilibriums and the convergence of best response dynamics when players 
compete on multiple servers or links simultaneously \cite{wu,feldman}.

\emph{Brief Summary:} Our study differs from the literature on timeline competition and contest theory in two aspects: 
i) we consider the presence of a misbehaving user that may appear in targeted advertising or security game, 
while all the users 
are assumed to be selfish in almost all the related works; ii) we study the worst and the best performance 
of the contest with heterogeneous users that have not been considered before. 
Our study distinguishes from the literature on bandwidth allocation games in three 
aspects: iii) we consider finite number of users; 
iv) three different performance measures are compared and v) both the lower and (some) upper bounds are investigated.

\section{Conclusion}
\label{sec:conclusion} 

This work studies the efficiency of the price-anticipating Kelly mechanism in a generalized system consisting of 
a finite number of benign users and a misbehaving user. 
We model their competition as a non-zero sum game and then investigate how the misbehaving user influences the 
three imporant metrics on the price-anticipating Kelly mechanism: 
the total utility and the total net utility of benign users as well as the network operator's revenue at the NE. 
We compare each of them in the scenarios whether the misbehaving user exists or not in the game. We obtain 
the lower bounds for the general utility functions, and the upper bounds for the linear utility functions. 
Both the upper and the lower bounds are found in the regime of finite number of benign users, while 
only the lower bound of only one metric has been analyzed with infinite number of benign users in the literature. 
Our study reveals two interesting observations on the misbehaving behaviors.  
Firstly, there exist certain ranges for the misbehaving user's willingness factor upon which 
a slight increase in the willingness factor can reduce the utility and the net utility 
of benign users remarkably. Secondly, the misbehaving user 
always brings more revenue to the network operator. When the willingness factor is large, 
the network operator's revenue at the NE can be even greater than the maximum revenue he receives 
in the absence of the misbehaving user. 

\bibliographystyle{IEEEtran}
\bibliography{IEEEabrv,mybibfile}




\clearpage
\section*{Appendix: Proofs}

\subsection*{Proof of Theorem \ref{theorem:linear_ne}}
When the utility functions are linear, the NE conditions are expressed as below:
\begin{eqnarray}
&&\frac{v_i z_{-i}^*}{(z^*)^2} -1  \left\{\begin{matrix}
= 0   &\textrm{ if }  x_i^* > 0\\
\leq 0 &\textrm{ if }  x_i^* = 0 
\end{matrix}\right. \quad \forall i\geq 1,
\label{eq:proof_theorem2_no1}\\
&&\theta \sum\nolimits_{i{=}1}^{N} \frac{v_ix_{i}^*}{(z^*)^2} - 1\left\{\begin{matrix}
=0   &\textrm{ if }  x_0^* > 0 \\
\leq 0 &\textrm{ if }   x_0^* = 0 
\end{matrix}\right..
\label{eq:proof_theorem2_no2}
\end{eqnarray}
We first show that if $n$ out of $N$ benign users participate in the NE, they must be 
$\SA_1$, $\SA_2$, $\cdots$, $\SA_n$. This is easy to prove by contradiction. 
Consider two benign users $\SA_{i}$ and $\SA_{j}$ with $v_i>v_j$, $x_i^*=0$ and $x_j^*>0$. 
The above NE conditions do not hold for such $\SA_i$ and $\SA_j$. This property paves the way of 
searching the NE step-by-step. 

i) When only $n$ benign users participate in the NE, the NE conditions give rise to the following equation:
\begin{eqnarray}
\theta &<& \frac{(z^*)^2}{\sum_{i=1}^nv_ix_i^*} \nonumber\\
&=& \frac{n-1}{\sum\nolimits_{j=1}^{n}v_j\sum\nolimits_{j=1}^{n}1/v_j {-}n(n{-}1)}. 
\end{eqnarray}
This inequality serves as the sufficient condition of excluding the misbehaving user at the NE. 
Summing equations in \eqref{eq:proof_theorem2_no1} for all $i\in[1, n]$, we obtain the total bid of all
benign users and then compute each bid $x_i^*$ by
\begin{eqnarray}
\left\{\begin{matrix}
x_i^* = \big(\frac{\sum\nolimits_{j=1}^{n}1/v_j}{n{-}1} {-} \frac{1}{v_i}
\big)\big(\frac{n{-}1}{\sum\nolimits_{j=1}^{n}1/v_j}\big)^2  \\
x_0^*=0  
\end{matrix}\right.
\label{eq:theorem2_thetacondition}
\end{eqnarray}

ii) When both $\SA_0$ and $n$ benign users participate in the NE, we sum the NE conditions of 
$n$ benign users and that of $\SA_0$ to compute $z^*$ first, and then solve all $x_i^*$, 
\begin{eqnarray}
\left\{\begin{matrix}
x_i^* = \big(\frac{n+1/\theta}{\sum\nolimits_{j=1}^{n}v_j} {-} \frac{1}{v_i}
\big)\big(\frac{\sum\nolimits_{j=1}^{n}v_j}{n{+}1/\theta}\big)^2 \\
x_0^*=\frac{\sum\nolimits_{j=1}^{n}v_j}{n+1/\theta}
\big( \frac{\sum\nolimits_{j=1}^{n}v_j\sum\nolimits_{j=1}^{n}1/v_j}{n+1/\theta}{+} 1 {-} n\big)
\end{matrix}\right.;
\end{eqnarray}

iii) The NE conditions also require $x_i^* > 0$ for $1\leq i\leq n$ at the NE. 
If we search from $n=N$, the NE will be found in no larger than $N$ steps, which is very intuitive. 

iv). When $x_0^* > 0$, the utility of benign users is given by 
\begin{eqnarray}
\SU = \frac{\sum\nolimits_{i=1}^{n}v_i}{n\theta+1} \nonumber
\end{eqnarray}
which is strictly decreasing with regard to $\theta$.
The net utility of benign users is expressed as 
\begin{eqnarray}
\SV = \frac{(1-n\theta)\sum\nolimits_{i=1}^{n}v_i}{n\theta+1}  + \sum\nolimits_{i=1}^{n}\frac{1}{v_i}
\frac{\theta^2(\sum\nolimits_{i=1}^{n}v_i)^2}{(1+n\theta)^2}.
\label{eq:netutility_theorem2}
\end{eqnarray}
Taking the derivative of $\SV$ over $\theta$, we obtain 
\begin{eqnarray}
\frac{d\SV}{d\theta} = \frac{2\sum\nolimits_{i=1}^{n}v_i(\sum_{i=1}^{n}\frac{1}{v_i}\sum_{i=1}^{n}v_i\theta -n(n\theta+1))}{(n\theta+1)^3}.
\label{eq:netutilityderivative_theorem2}
\end{eqnarray}
Given the condition of \eqref{eq:theorem2_thetacondition}, the above derivative is below 0. 
Hence, the net utility of benign users is also a decreasing function of $\theta$ when $\SA_0$ 
participates in the NE. \done

\subsection*{Proof of Theorem \ref{theorem:no2_participation}}

i). The first item can be derived directly from the participation condition of the misbehaving user
\begin{eqnarray}
\theta > \frac{n-1}{\sum\nolimits_{i=1}^{n}v_i\sum\nolimits_{i=1}^{n}1/v_i - n(n-1)}.
\end{eqnarray}
The expression $\sum\nolimits_{i=1}^{n}v_i\sum\nolimits_{i=1}^{n}1/v_i$ has the minimum value $n^2$ 
for $0\leq v_i \leq 1$ $i=1,\cdots, N$. Thus, the participation threshold of the misbehaving user 
is equal to $\frac{n{-}1}{n}$, given $n$ benign users send positive bids to the network operator. 
Since there has $\frac{n{-}1}{n}<\frac{N{-}1}{N}$, according to \eqref{eq:linearutility_no1}, 
$x_0^*$ is always positive at the NE no matter how many benign users participate for $\theta > \frac{N-1}{N}$.

ii). For the second item, if $x_i^* > 0$ and $x_j^* > 0$, the NE conditions yield the following equality
\begin{eqnarray}
\frac{(x_i^*-x_j^*)}{(\sum\nolimits_{i=1}^N x_{i}^*+ x_0^*)^2} = \frac{1}{v_j}-\frac{1}{v_i}.
\end{eqnarray} 
If $v_i>v_j$, there must have $x_i^*>x_j^*$. 

iii) For the last item, we prove it by contradiction. We assume $x_j^* = 0$ and $x_i^* > 0$ at 
the NE when the valuations of all the benign users are identical. Then 
the expression \eqref{eq:onetarget_opt1} gives rise to
\begin{eqnarray}
\frac{1}{\sum\nolimits_{i=1}^N x_{i}^*+ x_0^*} \leq \frac{1}{v_j}
\end{eqnarray}
and
\begin{eqnarray}
\frac{1}{\sum\nolimits_{i=1}^N x_{i}^*+ x_0^*} - \frac{x_i^*}{(\sum\nolimits_{i=1}^N x_{i}^*+ x_0^*)^2} = \frac{1}{v_i}.
\end{eqnarray}
Because $v_i$ equals to $v_j$, it is easy to conclude
\begin{eqnarray}
\frac{x_i^*}{(\sum\nolimits_{i=1}^N x_{i}^*+ x_0^*)^2} \leq 0,
\end{eqnarray}
which is not true. Thus, $x_i^*$ must be positive for all the benign users at the NE. \done

\subsection*{Proof of Theorem \ref{theorem:all_lowerbounds_linear}}

(1) \textit{\textbf{B1:} (NE Utility over Maximum)} 

We consider two cases at the NE separately, i.e. $x_0^*=0$ and $x_0^*>0$. 

\noindent \textit{Case 1: $x_0^*=0$.} When the misbehaving user does not participate at the NE, there are 
at least two benign users with positive bids. 
We assume that the top $n$ benign users participate at the NE. 
Then, the minimum total utility is the 
result of the following problem:
\begin{eqnarray}
\min && \SU(\mathbf{d^*})  \\
\textrm{s.t.} && 0\leq v_i \leq 1,\quad \forall i=2\cdots,n.
\end{eqnarray}
When $\SA_0$ does not participate, the NE conditions are expressed as
\begin{eqnarray}
\frac{v_i}{\sum\nolimits_{j=1}^{n}x_j^*} - \frac{v_ix_i^*}{(\sum\nolimits_{j=1}^{n}x_j^*)^2} = 1, \quad \forall i=2\cdots,n.
\end{eqnarray}
After some simple manipulations, there has
\begin{eqnarray}
\sum\nolimits_{i=1}^{n}x_j^* = \frac{n-1}{\sum\nolimits_{j=1}^n1/v_j}.
\end{eqnarray}
The utility of each $\SA_i$ is obtained by
\begin{eqnarray}
v_id_i^* = v_i - \sum\nolimits_{j=1}^{n}x_j^* = v_i - \frac{n-1}{\sum\nolimits_{j=1}^n1/v_j}.
\label{eq:theorem4_proof_no1}
\end{eqnarray}
We next show that $\SU$ is a convex function over the vector of valuations $\{v_i\}_{i=2}^n$
and the minimum is obtained at the boundary of the feasible region. 
The Hessian matrix of $\SU$ is given by 
\begin{eqnarray}
\SH = \frac{2n(n{-}1)}{(\sum\nolimits_{i=1}^{n}1/v_i)^3}\cdot
\begin{bmatrix}
\sum_{i=1,\neq 2}^{n}\frac{1}{v_i} & -\frac{v_2}{v_3^2} & \cdots & -\frac{v_2}{v_n^2}\\
-\frac{v_3}{v_2^2} & \sum_{i=1,\neq 3}^{n}\frac{1}{v_i} & \cdots & -\frac{v_3}{v_n^2} \\
\cdots & \cdots & \cdots & \cdots \\
-\frac{v_n}{v_2^2} & -\frac{v_n}{v_3^2} & \cdots & \sum_{i=1}^{n{-}1}\frac{1}{v_i} 
\end{bmatrix}.\nonumber
\end{eqnarray}
Here, we omit the positive constant $ \frac{2n(n{-}1)}{(\sum\nolimits_{i=1}^{n}1/v_i)^3}$ 
and denote the Hessian matrix as $\hat{\SH}$.
Define two matrices $\hat{\SH}_1$ and $\hat{\SH}_2$ as
\begin{eqnarray}
\hat{\SH}_1 = \textbf{Diag} (\sum\nolimits_{i=1}^{n}\frac{1}{v_i}),\quad
\hat{\SH}_2 = 
\begin{bmatrix}
\frac{v_2}{v_2^2} & \frac{v_2}{v_3^2} & \cdots & \frac{v_2}{v_n^2}\\
\frac{v_3}{v_2^2} & \frac{v_3}{v_3^2} & \cdots & \frac{v_3}{v_n^2} \\
\cdots & \cdots & \cdots & \cdots \\
\frac{v_n}{v_2^2} & \frac{v_n}{v_3^2} & \cdots & \frac{v_n}{v_n^2}
\end{bmatrix}\nonumber
\end{eqnarray}
where $\hat{\SH} = \hat{\SH}_1 - \hat{\SH}_2$. $\hat{\SH}_1$ is a full-rank diagonal 
matrix whose diagonal elements are the same, while the rank of $\hat{\SH}_2$ is only 1. 
Thus, the matrix $\hat{\SH}$ has $m-1$ identical eigenvalue which is 
$\sum\nolimits_{i=1}^{n}\frac{1}{v_i}$. The last unknown eigenvalue is computed as $\frac{1}{v_1}$ and we 
validate it as the following. Define a new matrix as $\hat{\SH}_3$ that has
\begin{eqnarray}
\hat{\SH}_3 &=& \hat{\SH} - \textbf{Diag}(\frac{1}{v_1}) \nonumber\\
&=&
\begin{bmatrix}
\sum_{i=3}^{n}\frac{1}{v_i} & -\frac{v_2}{v_3^2} & \cdots & -\frac{v_2}{v_n^2}\\
-\frac{v_3}{v_2^2} & \sum_{i=2,\neq 3}^{n}\frac{1}{v_i} & \cdots & -\frac{v_3}{v_n^2} \\
\cdots & \cdots & \cdots & \cdots \\
-\frac{v_n}{v_2^2} & -\frac{v_n}{v_3^2} & \cdots & \sum_{i=2}^{n-1}\frac{1}{v_i}
\end{bmatrix}\nonumber.
\end{eqnarray}
For each column $j (j\geq 2)$, we multiply it by $\frac{v_{j{+}1}}{v_2}$ and add to column 1. Then, 
column 1 becomes a zero vector, which validates the existence of an eigenvalue to be $\frac{1}{v_1}$. 
All the eigenvalues of the Hessian matrix $\SH$ are positive which means that $\SH$ is strictly 
positive definite. Hence, $\SU$ is a convex function so that it is maximized at the
point $v_i = 1$ for $1\leq i \leq n$ and the maximum value is 1 (i.e. $\SU_{max} = 1$).

Because $\SU$ is strictly convex over the set of valuations $\{v_i\}_{i=2}^{n}$, 
the minimum utility $\SU$ is chosen at the point that has $\frac{\partial \SU}
{\partial v_i} = 0$ for all $2\leq i \leq n$ if they exist. 
By letting the first-order derivatives be 0, we obtain the optimal valuation $v_i^*$ by
\begin{eqnarray}
v_i^* = \sqrt{n(n{-}1)} - (n{-}1), \quad 2\leq i \leq n.
\label{eq:theorem4_proof_no2}
\end{eqnarray}
Submitting Eq.\eqref{eq:theorem4_proof_no2} to \eqref{eq:theorem4_proof_no1},
we obtain the lowest total utility of benign users by
\begin{eqnarray}
\SU(\mathbf{d}^*) = 1 - (n{-}1) \big(  \sqrt{n} - \sqrt{n-1}\big)^2.
\label{eq:theorem4_proof_no3}
\end{eqnarray}
To find the monotonicity between $\SU$ and $n$, 
we assume that $n$ is a
continuous variable. We take the derivative of $\SU$ over $n$ and obtain
\begin{eqnarray}
\frac{d\SU}{dn} \!\!&=&\!\! 3\sqrt{n(n-1)}{-}3(n{-}1) {-} n {+}(n{-}1)\sqrt{\frac{n{-}1}{n}} \nonumber\\
\!\!&=&\!\! \frac{1}{\sqrt{n}}(\sqrt{n-1} - \sqrt{n})^3 < 0.
\label{eq:theorem4_proof_no4}
\end{eqnarray}
Given $N$ benign users competing for the resources, the minimum total utility is obtained by 
\begin{eqnarray}
\SU_{mal} \geq 1 - (N{-}1)  \big(  \sqrt{N} - \sqrt{N-1}\big)^2.
\label{eq:theorem4_proof_no5}
\end{eqnarray}
The total bid of the benign users at this NE is computed as
\begin{eqnarray}
\sum\nolimits_{i=1}^{N}x_i^* = \frac{N-1}{\sum\nolimits_{j=1}^{N}1/v_j} = \frac{N-1}{\sqrt{N(N{-}1)}+N}.
\label{eq:theorem4_proof_no6}
\end{eqnarray}
Since the misbehaving user does not participate, the following inequality holds
\begin{eqnarray}
\theta \leq \frac{\sum\nolimits_{i=1}^{N}x_i^*}{\SU(N)} = \frac{\frac{N-1}{\sqrt{N(N{-}1)}+N}}{1-(N{-}1) ( \sqrt{N} - \sqrt{N{-}1})^2} 
\label{eq:theorem4_proof_no7}
\end{eqnarray}
according to the NE conditions in \eqref{eq:onetarget_opt2}. As $N$ approaches infinity,
the asymptotic total utility is given by
\begin{eqnarray}
\lim_{N\rightarrow \infty}\SU(\mathbf{d}^*) \!\!&=&\!\! 1 - \lim_{N\rightarrow \infty}(N{-}1) (  \sqrt{N} - \sqrt{N-1})^2 \nonumber\\
\!\!&=&\!\! 1- \lim_{N\rightarrow \infty}\frac{N{-}1}{(\sqrt{N} + \sqrt{N-1})^2} \nonumber\\
\!\!&=&\!\! 3/4.
\label{eq:theorem4_proof_no8}
\end{eqnarray}


\noindent \textit{Case 2: $x_0^*>0$.} The NE conditions result in the following equations
\begin{eqnarray}
&&\frac{v_i}{\sum\nolimits_{j=0}^{n}x_j^*} - \frac{v_ix_i^*}{(\sum\nolimits_{j=0}^{n}x_j^*)^2} = 1, 
\quad \forall i=1,\cdots, n, \\
&&\frac{\sum\nolimits_{j=1}^{n}v_ix_i^*}{(\sum\nolimits_{j=0}^{n}x_j^*)^2} = \frac{1}{\theta}. 
\end{eqnarray}
Summing the above equations together, we have 
\begin{eqnarray}
\sum\nolimits_{j=0}^{n}x_j^* = \frac{\theta\sum\nolimits_{j=1}^{n}v_j}{n\theta+1}
\end{eqnarray}
and
\begin{eqnarray}
\SU(\mathbf{d}^*) = \frac{\sum\nolimits_{j=1}^{n}v_j}{n\theta+1}.
\label{eq:proof_totutility_withA0}
\end{eqnarray}
Denote by $\underline{v}$ the minimum valuation of $\SA_2$ untill $\SA_{n}$. 
Here, $\underline{v}$ must guarantee that each of these users
pays a non-zero bid at the NE. Thus, 
the worst total utility of benign users is expressed as
\begin{eqnarray}
\SU(\mathbf{d}^*) = \frac{1+(n{-}1)\underline{v}}{n\theta+1}.
\label{eq:theorem4_proof_no11}
\end{eqnarray}
The NE condition in \eqref{eq:onetarget_opt2} requires
\begin{eqnarray}
\underline{v}(n+1/\theta) \geq 1 + (n-1)\underline{v}.
\label{eq:theorem4_proof_no12}
\end{eqnarray}
The above inequality gives rise to
\begin{eqnarray}
\underline{v} \geq \theta/(1+\theta).
\label{eq:theorem4_proof_no13}
\end{eqnarray}
Submitting inequality \eqref{eq:theorem4_proof_no13} to \eqref{eq:theorem4_proof_no11},
we obtain
\begin{eqnarray}
\SU_{mal} \geq \frac{1}{1+\theta}.
\label{eq:theorem4_proof_no14}
\end{eqnarray}
The set of valuations $\{v_i\}_{i=2}^{N}$ to minimize $\SU$ are not unique.
For instance, $\SU$ is minimized in the scenario where 
only two players, $\SA_0$ and $\SA_1$, participate in the competition at the NE, while 
the valuations of other benign users can be chosen below $\underline{v}$ arbitrarily. 

Combing the lower bounds in both cases, we can conclude that the utility is lower bounded by
\begin{eqnarray}
\SU_{mal} \geq \{1{-}(N{-}1)(\sqrt{N}{-}\sqrt{N{-}1})^2, (1+\theta)^{-1}\} \nonumber
\end{eqnarray}
where the crosspoint is taken at 
\begin{eqnarray}
\theta = \big((\sqrt{\frac{N}{N{-}1}}{+}1)^2 {-} 1\big)^{{-}1}.\nonumber
\end{eqnarray}

(2) \textit{\textbf{B2:} (NE Utility With/Without Misbehaving User)}

This theorem compares the NEs of two games, $\mathbf{G_A}$ and $\mathbf{G_B}$: 
the first one excludes the misbehaving user, while the second one considers the possible 
participation of the misbehaving user. We suppose that the participation of $n_A$ benign users 
at the NE in the former and the participation of $n_B$ benign 
users at the NE in the latter, given the set of valuations $\{v_i\}_{i=1}^{N}$. 
In these two games, $n_A$ and $n_B$ might not be the same. We prove this theorem via four steps.

\textit{Step 1: Proving $n_A \geq n_B$}. When the misbehaving user participates, 
the number of benign users that pay non-negative bids may decrease. 
This step can be easily proved by contradiction from the NE conditions.

\textit{Step 2: Approximating the ratio of two utilities}. The ratio $\frac{\SU_{mal}}{\SU_{nom}}$ is computed by
\begin{eqnarray}
\frac{\SU_{mal}}{\SU_{nom}} = \frac{\sum\nolimits_{i=1}^{n_B}v_i}{1+n_B\theta}\cdot \big(\sum\nolimits_{i=1}^{n_A}v_i - \frac{n_A(n_A-1)}{\sum\nolimits_{i=1}^{n_A}1/v_i}\big)^{-1}.
\label{eq:SU_malvsnom_no1}
\end{eqnarray}

\noindent According to the NE conditions of the game $\mathbf{G_B}$, since there are $n_B$ benign users participating 
in the competition, we have 
\begin{eqnarray}
\frac{\sum\nolimits_{i=1}^{n_B}v_i}{\sum\nolimits_{i=1}^{n_B}x_i^*} = n_B+\frac{1}{\theta}   \quad \textrm{and} \quad x_i^* = 0 \;\;\;\forall i>n_B.
\label{eq:SU_malvsnom_no2}
\end{eqnarray}
For any $n$ with $n>n_B$, the following inequality holds
\begin{eqnarray}
\frac{v_n}{\sum\nolimits_{i=1}^{n}x_i^*} = \frac{v_n}{\sum\nolimits_{i=1}^{n_B}x_i^*}  \leq 1
\label{eq:SU_malvsnom_no3}
\end{eqnarray}
due to $x_i^* = 0$. Hence, for $n_A$ and $n_B$ with $n_A\geq n_B$, we obtain 
\begin{eqnarray}
\frac{\sum\nolimits_{i=1}^{n_A}v_i}{\sum\nolimits_{i=1}^{n_A}x_i^*} \!\!\!&=&\!\!\! \frac{\sum\nolimits_{i=1}^{n_A}v_i}{\sum\nolimits_{i=1}^{n_B}x_i^*}  
=\frac{\sum\nolimits_{i=1}^{n_B}v_i}{\sum\nolimits_{i=1}^{n_B}x_i^*} +  \frac{\sum\nolimits_{i=n_B+1}^{n_A}v_i}{\sum\nolimits_{i=1}^{n_B}x_i^*}
\nonumber \\
\!\!\!&\leq&\!\!\! n_B+\frac{1}{\theta} + (n_A - n_B) = n_A + \frac{1}{\theta}
\label{eq:SU_malvsnom_no4}
\end{eqnarray}
which implies
\begin{eqnarray}
\frac{\sum\nolimits_{i=1}^{n_A}v_i}{n_A+\frac{1}{\theta}}  \leq \frac{\sum\nolimits_{i=1}^{n_B}v_i}{n_B+\frac{1}{\theta}} .
\label{eq:SU_malvsnom_no5}
\end{eqnarray}
Submitting \eqref{eq:SU_malvsnom_no5} to \eqref{eq:SU_malvsnom_no1}, we derive the approximated lower bound by 
\begin{eqnarray}
\frac{\SU_{mal}}{\SU_{nom}} \geq \frac{\sum\nolimits_{i=1}^{n_A}v_i}{1+n_A\theta}\cdot \big(\sum\nolimits_{i=1}^{n_A}v_i - \frac{n_A(n_A-1)}{\sum\nolimits_{i=1}^{n_A}1/v_i}\big)^{-1}.
\label{eq:SU_malvsnom_no6}
\end{eqnarray}
Note that the NE of the game $\mathbf{G_A}$ must involve two benign users. 
Eq.\eqref{eq:SU_malvsnom_no6} serves as a lower bound for $n_A\geq 2$. 
When only $\SA_1$ participates in the NE of the game $\mathbf{G_B}$, the utility of $\SA_1$ is given by 
$\SU_{mal} = \frac{1}{1+\theta}$. The total utility of benign users at the NE of the game $\mathbf{G_A}$ 
is no larger than 1. In this scenario, Eq.\eqref{eq:SU_malvsnom_no6} still holds at $n_A=1$.
Therefore, Eq.\eqref{eq:SU_malvsnom_no6} captures the lower bound of $\frac{\SU_{mal}}{\SU_{nom}}$ 
for any $n_A\in[1, N]$. We drop all the subscripts and obtain
\begin{eqnarray}
\frac{\SU_{mal}}{\SU_{nom}} \geq \frac{\sum\nolimits_{i=1}^{n}v_i\sum\nolimits_{i=1}^{n}1/v_i}
{\sum\nolimits_{i=1}^{n}v_i\sum\nolimits_{i=1}^{n}1/v_i - n(n-1)} \cdot \frac{1}{1+n\theta}, \; 1{\leq} n{\leq} N. 
\label{eq:SU_malvsnom_no7}
\end{eqnarray}
We let $\Delta$ denote the expression $\sum\nolimits_{i=1}^{n}\!v_i\sum\nolimits_{i=1}^{n}\!\frac{1}{v_i}$. 
The derivative of the right-hand of Eq.\eqref{eq:SU_malvsnom_no7} over $\Delta$ yields
\begin{eqnarray}
\frac{-n(n-1)}
{(\Delta- n(n-1))^2} \cdot \frac{1}{1+n\theta} \leq 0
\label{eq:SU_malvsnom_no8}
\end{eqnarray}
where the equality holds only with $n=1$. Therefore, the lower bound of $\frac{\SU_{mal}}{\SU_{nom}}$ satisfies 
\begin{eqnarray}
\frac{\SU_{mal}}{\SU_{nom}} \geq \frac{\max{\Delta}}
{\max{\Delta} - n(n-1)} \cdot \frac{1}{1+n\theta}. 
\label{eq:SU_malvsnom_no9}
\end{eqnarray}

\textit{Step 3: Finding the maximum of the expression $\sum\nolimits_{i=1}^{n}\!v_i\sum\nolimits_{i=1}^{n}\!\frac{1}{v_i}$}. 

To solve $\max \Delta $, we first analyze its first-order derivatives over each $v_i$ for $1\leq i \leq n$. 
Since $v_n$ is no larger than any other $v_i$, there exists
\begin{eqnarray}
\frac{d\Delta}{dv_n} = \sum_{i=1}^{n}\frac{1}{v_i} - \frac{1}{v_n^2} \sum_{i=1}^{n}v_i    =
\sum_{i=1}^{n}\frac{v_n^2 - v_i^2}{v_i v_n^2} \leq 0.
\label{eq:SU_malvsnom_no10}
\end{eqnarray}
The equality holds only when all $v_n$ are the same. Hence, $\Delta$ is a decreasing function of $v_n$. 
The maximum $\Delta$ is obtained at the point that $v_n$ reduces to its minimum value. When $v_n$ further decreases, 
$\SA_n$ will not participate in the competition. For ease of notation, we denote $\underline{v}$ as 
the minimum value for $v_n$. 

We next analyze the optimal selection of $v_i$ for $2\leq i \leq (n{-}1)$. By taking the first and the second-order 
derivatives, we obtain the following equations
\begin{eqnarray}
\frac{d\Delta}{dv_i} \!\!&=&\!\! \sum_{j=1}^{n}\frac{1}{v_j} - \frac{1}{v_i^2} \sum_{j=1}^{n}v_j ,
\label{eq:SU_malvsnom_no11}\\
\frac{d^2\Delta}{dv_i^2} \!\!&=&\!\! \frac{2}{v_i^3} \sum_{j{=}1,{\neq i}}^{n} v_j > 0
\label{eq:SU_malvsnom_no12}
\end{eqnarray}
for all $2\leq i\leq (n{-}1)$. By letting $\frac{d\Delta}{dv_i} =0$, we 
can obtain a set of $\{v_i\}_{i=2}^n$ to reach the extremum of $\Delta$. 
However, their second-order derivatives are strictly positive.  
This means that the maximum of $\Delta$ is not reached at these $\{v_i\}_{i=2}^n$, but at the boundaries, i.e. $v_i = 1$ 
or $v_i = \underline{v}$ for all $2\leq i\leq (n{-}1)$. 

We suppose $v_i = 1$ for $1\leq i \leq k$ and $v_i = \underline{v}$ for $k{+}1\leq i \leq n$. 
Then, there yields
\begin{eqnarray}
\Delta \!\!&=&\!\! (k + (n-k)\underline{v})(k + \frac{n-k}{\underline{v}}) \nonumber\\
\!\!&=&\!\! k^2 + (n-k)^2 + k(n-k)(\underline{v}+\frac{1}{\underline{v}}) \nonumber\\
\!\!&=&\!\! n^2 + k(n-k)(\underline{v}+\frac{1}{\underline{v}}-2)  \nonumber\\
\!\!&=&\!\! n^2 +k(n-k) (\sqrt{\underline{v}} - \sqrt{\frac{1}{\underline{v}}})^2 \nonumber\\
\!\!&=&\!\! n^2 +(\frac{n^2}{4} - (\frac{n}{2}-k)^2 )(\sqrt{\underline{v}} - \sqrt{\frac{1}{\underline{v}}})^2 .
\label{eq:SU_malvsnom_no13}
\end{eqnarray}
Given $n$ and $\sqrt{\underline{v}}$, we find that the maximum $\Delta$ is obtained at 
$k=\frac{n}{2}$ if $n$ is an even number, and is obtained at $k=\lceil \frac{n}{2}\rceil$ or $k=\lfloor \frac{n}{2}\rfloor$
if $n$ is an odd number. Although $k=\lceil \frac{n}{2}\rceil$ and $k = \lfloor \frac{n}{2}\rfloor$ give rise to 
the same $\Delta$ for fixed $\underline{v}$, these two scenarios correspond to different $\underline{v}$ at the NE. 
In what follows, we examine the choice of $\underline{v}$ from the NE conditions. 

Case i): $n$ is an even number. Here, $\underline{v}$ is reached at the boundary that $\SA_n$ pays an 
arbitrarily small amount, i.e. $x_n^*$ is infinitely approaching 0. The NE conditions result in the following equations
\begin{eqnarray}
\frac{\sum\nolimits_{i=1}^{n}v_i}{\sum\nolimits_{i=1}^{n}x_i^*} = n+\frac{1}{\theta} \quad \textrm {and} \quad 
\frac{v_n}{\sum\nolimits_{i=1}^{n}x_i^*} = \frac{v_n}{\sum\nolimits_{i=1}^{n}x_i^*} = 1. 
\label{eq:SU_malvsnom_no14}
\end{eqnarray}
The above equations solve $\underline{v}$ by 
\begin{eqnarray}
\underline{v} = \frac{n\theta}{n\theta + 2}
\label{eq:SU_malvsnom_no15}
\end{eqnarray}
The maximum $\Delta$ is thus computed as
\begin{eqnarray}
\max_{even}\Delta \!\!&=&\!\! \frac{n^2}{4}(1+\underline{v})(1+\frac{1}{\underline{v}}) \nonumber\\
\!\!&=&\!\! n^2+\frac{n}{\theta} - \frac{n^2}{n\theta+2} - \frac{n}{n\theta^2+2\theta}. 
\label{eq:SU_malvsnom_no16}
\end{eqnarray}

Case ii): $n$ is an odd number and $k = \lfloor \frac{n}{2}\rfloor$. Eq.\eqref{eq:SU_malvsnom_no14} gives rise to 
\begin{eqnarray}
\lfloor \frac{n}{2}\rfloor + \lceil \frac{n}{2}\rceil \underline{v} = (n+\frac{1}{\theta})\underline{v}. 
\label{eq:SU_malvsnom_no17}
\end{eqnarray}
Hence, $\underline{v}$ is given by 
\begin{eqnarray}
\underline{v}  =  \frac{\lfloor \frac{n}{2}\rfloor}{\lfloor \frac{n}{2}\rfloor + \frac{1}{\theta}}.
\label{eq:SU_malvsnom_no18}
\end{eqnarray}

Case iii): $n$ is an odd number and $k = \lceil \frac{n}{2}\rceil$. Following the same technique, we have 
\begin{eqnarray}
\underline{v}  =  \frac{\lceil \frac{n}{2}\rceil}{\lceil \frac{n}{2}\rceil + \frac{1}{\theta}} >
\frac{\lfloor \frac{n}{2}\rfloor}{\lfloor \frac{n}{2}\rfloor + \frac{1}{\theta}}.
\label{eq:SU_malvsnom_no19}
\end{eqnarray}

Obviously, $\Delta$ is a decreasing function of $\underline{v}$ according to Eq.\eqref{eq:SU_malvsnom_no13}. 
Therefore, we only consider Case i) and Case ii) to compute the lower bound of $\frac{\SU_{mal}}{\SU_{mal}}$.
The maximum $\Delta$ at the case of odd $n$ is given by
\begin{eqnarray}
\max_{odd}\Delta\!\!\!\!\!&=&\!\!\!\!\! (\lfloor \frac{n}{2}\rfloor + \lceil \frac{n}{2}\rceil \underline{v})
(\lfloor \frac{n}{2}\rfloor + \lceil \frac{n}{2}\rceil \frac{1}{\underline{v}}) \nonumber\\
\!\!\!\!\!&=&\!\!\!\!\!n^2 {+} 
\frac{n(n{+}1)}{\theta(n{-}1)} {-}\frac{n(n{+}1)}{\theta(n{-}1){+}2} {-}\frac{(n{+}1)^2}{\theta^2(n{-}1)^2 {+} 2\theta(n{-}1)}. 
\label{eq:SU_malvsnom_no20}
\end{eqnarray}

\textit{Step 4: Analyzing the lower bound of $\frac{\SU_{mal}}{\SU_{mal}}$}. 

We consider the two cases with regard to $n$ in \emph{Step 3} separately. 

Case i): $n$ is an even number. Submitting Eq.\eqref{eq:SU_malvsnom_no16} to Eq.\eqref{eq:SU_malvsnom_no9},
we obtain the following inequality 
\begin{eqnarray}
\frac{\SU_{mal}}{\SU_{nom}} \geq \frac{1+n\theta}{1+2\theta+n\theta^2}.
\label{eq:SU_malvsnom_no21}
\end{eqnarray}
Then, there exists
\begin{eqnarray}
\frac{\SU_{mal}}{\SU_{nom}} - \frac{1}{1+\theta}\geq \frac{(n-1)\theta}{(1+2\theta+n\theta^2)(1+\theta)} >0
\label{eq:SU_malvsnom_no22}
\end{eqnarray}
where $n$ is no less than 2. Hence, the lower bound of $\frac{\SU_{mal}}{\SU_{nom}}$ is expressed as 
\begin{eqnarray}
\frac{\SU_{mal}}{\SU_{nom}} > \frac{1}{1+\theta}.
\label{eq:SU_malvsnom_no23}
\end{eqnarray}

Case ii): $n$ is an odd number and $k = \lfloor \frac{n}{2}\rfloor$. Submitting Eq.\eqref{eq:SU_malvsnom_no18}
to Eq.\eqref{eq:SU_malvsnom_no9}, the lower bound of $\frac{\SU_{mal}}{\SU_{nom}}$ is given by 
\begin{eqnarray}
\frac{\SU_{mal}}{\SU_{nom}} \geq \frac{n(n-1)\theta + (n+1)}{(n+1)+n(n-1)\theta^2 +2\theta n}.
\label{eq:SU_malvsnom_no24}
\end{eqnarray}
Then, we derive the following inequality
\begin{eqnarray}
\frac{\SU_{mal}}{\SU_{nom}} - \frac{1}{1+\theta} \geq \frac{\theta(n-1)^2}{((n+1)+n(n-1)\theta^2 +2\theta n)(1{+}\theta)} \geq 0.
\label{eq:SU_malvsnom_no25}
\end{eqnarray} 
Therefore, the lower bound of $\frac{\SU_{mal}}{\SU_{nom}}$ is formally given by 
\begin{eqnarray}
\frac{\SU_{mal}}{\SU_{nom}} \geq \frac{1}{1+\theta}
\label{eq:SU_malvsnom_no26}
\end{eqnarray} 
where it is tight for $n=1$.

(3) \textit{\textbf{B3:} (NE Net Utility over Maximum)}

First of all, the social optimal net utility $\SV_{max}$ approaches 1 asymptotically since $\SU_{max}$ 
is 1. We suppose  $x_i^*>0$ for $1\leq i\leq n$ and $x_i^*=0$ for 
$n{+}1\leq i\leq N$ at the NE. Similar to the preceding proof, two cases, 
$x_0^*=0$ and $x_0^*>0$, are considered. 

\noindent \textit{Case 1: $x_0^*=0$.} The total net utility of benign users is obtained by 
\begin{eqnarray}
\SV(\mathbf{d}^*) = \sum\nolimits_{i=1}^{n}v_i - \frac{n^2-1}{\sum\nolimits_{i=1}^{n}1/v_i}. 
\label{eq:theorem5_proof_no1}
\end{eqnarray}
As proved before, $\SU(\mathbf{d}^*)$ is a strictly convex function over the set of valuations $\{v_i\}_{i=2}^{n}$, so is $\SV(\mathbf{d}^*)$. Hence, $\SV(\mathbf{d}^*)$ is minimized at its  
unique interior point if it exists at the feasible region. 
We differentiate $\SV(\mathbf{d}^*)$ over $v_i$ ($2\leq i \leq n$) and obtain
\begin{eqnarray}
\frac{\partial \SV}{\partial v_i} = 1- \frac{n^2{-}1}{(\sum\nolimits_{j=1}^{n}1/v_i)^2}\cdot \frac{1}{v_i^2}, \quad 2\leq i\leq n.
\label{eq:theorem5_proof_no2}
\end{eqnarray}
If there exists a feasible solution to $\frac{\partial \SU}{\partial v_i} = 0$
for all $2\leq i\leq n$, it minimizes $\SV(\mathbf{d}^*)$ with 
the participation of $n$ benign users. Due to the symmetric property
of Eq.\eqref{eq:theorem5_proof_no2}, all $v_i$  ($2\leq i\leq n$) are
the same when minimizing $\SV(\mathbf{d}^*)$. Hence, we can easily obtain
\begin{eqnarray}
v_a = v_i = \sqrt{n^2-1} - (n-1), \quad \forall \; 2\leq i\leq n.
\label{eq:theorem5_proof_no3}
\end{eqnarray}
Submitting Eq.\eqref{eq:theorem5_proof_no3} to \eqref{eq:theorem5_proof_no1},
we obtain
\begin{eqnarray}
\SV(\mathbf{d}^*) = 1+\frac{(n{-}1)(\sqrt{n^2-1}-(n{+}1))}{1+\frac{1}{2}\big(\sqrt{n^2{-}1}+(n{-}1)\big)}.
\label{eq:theorem5_proof_no4}
\end{eqnarray}
The total amount of bids generated by benign users at the NE is computed as
\begin{eqnarray}
\sum\nolimits_{i=1}^{n}x_i^* = \frac{n-1}{\sum\nolimits_{j=1}^{n}1/v_j} = \frac{n-1}{1+\frac{1}{2}\big(\sqrt{n^2{-}1}+(n{-}1)\big)}.
\label{eq:theorem5_proof_no5}
\end{eqnarray}
To enforce $x_0^* = 0$ at this NE, the willingness factor $\theta$ should satisfy
\begin{eqnarray}
\theta < \frac{(1+(n-1)v_a)(1+(n-1)/v_a)}{n-1} - n
\label{eq:theorem5_proof_no6}
\end{eqnarray}
according to the expression \eqref{eq:onetarget_opt2}. We next compare $\SV(\mathbf{d}^*)$
for different $n\in[2,N]$. Suppose that $n$ is a continuous variable. We
differentiate $\SV(\mathbf{d}^*)$ over $n$ and get $\frac{d\SV(\mathbf{d}^*)}{dn} <0$.
Therefore, the worst total net utility of benign users is obtained when all of them 
participate at the NE. The minimum total net utility is given by 
\begin{eqnarray}
\SV_{mal} \geq 1+\frac{(N{-}1)(\sqrt{N^2-1}-(N{+}1))}
{1+\frac{1}{2}\big(\sqrt{N^2{-}1}+(N{-}1)\big)}.
\label{eq:theorem5_proof_no6.1}
\end{eqnarray}

\noindent \textit{Case 2: $x_0^*>0$.} The total net utility of benign users is
expressed in the following equation
\begin{eqnarray}
\SV(\mathbf{d}^*) &=&  \SU(\mathbf{d}^*) - \sum\nolimits_{i=1}^{n}x_i^* \nonumber\\
&=&  \frac{1-n\theta}{1+n\theta} \sum_{i=1}^{n}v_i  + (\sum_{i=1}^{n}\frac{1}{v_i})\frac{(\theta\sum\nolimits_{i=1}^{n}v_i)^2}{(n\theta+1)^2}. 
\label{eq:theorem5_proof_no7}
\end{eqnarray}
We differentiate $\SV(\mathbf{d}^*)$ over $v_i$ and obtain
\begin{eqnarray}
\frac{d\SV}{dv_i} =\frac{1{-}n\theta}{1{+}n\theta}  + (\frac{\theta}{1{+}n\theta})^2(\frac{{-}1}{v_i^2}(\sum_{j=1}^{n}v_j)^2 {+} \sum_{j=1}^{n}\frac{2}{v_j}\sum_{j=1}^{n}v_j)
\label{eq:theorem5_proof_no8}
\end{eqnarray}
and
\begin{eqnarray}
\frac{d^2\SV}{dv_i^2} = (\frac{\theta}{1{+}n\theta})^2\frac{2}{v_i}\big((\frac{\sum_{j=1}^{n}v_j}{v_i}-1)^2 {+} \sum_{j=1,\neq i}^{n}\frac{1}{v_j}\big) >0.
\label{eq:theorem5_proof_no8_2}
\end{eqnarray}
From Eqs.\eqref{eq:theorem5_proof_no8} and \eqref{eq:theorem5_proof_no8_2}, 
there exists a unique interior point to minimize $\SV(\mathbf{d}^*)$ 
that satisfies $0\leq v_i \leq 1$ for all $i=2,\cdots,n$.
Due to the symmetric property, there have $v_i^*=v_j^*$ for any $i$ and $j$, $2\leq i,j\leq n$. 
Hence, we denote by $v_b^{(n)}$ the identical valuation of the benign users except $\SA_1$ at this interior point. 
By letting the derivative $\frac{d\SV}{dv_i}$ be 0, we obtain the following equation
\begin{eqnarray}
2(n-1) v^3 + (3-2n+\theta^{-2}) v^2 -1 = 0
\label{eq:theorem5_proof_no9}
\end{eqnarray}
where $v_b^{(n)}$ is the feasible solution. 
Note that the minimum might take values at the boundary if the interior point is outside of the feasible
region. Thus, we need to examine the possible scenarios where the valuations of some benign users 
are chosen at the boundary. 
i) $v_i^* = 1$ and $0< v_j^* <1$ for $2\leq i,j \leq n$. We will show that this scenario does not happen. For any $v_j^*$ with $0<v_j^*<1$, there exists $\frac{d\SV}{dv_j}|_{v_j = v_j^*} = 0$. 
Thus, there must have $\frac{d\SV}{dv_i}|_{v_i = 1} > 0$. 
This means that any local minimum cannot contain $v_i^* = 1$ for $2\leq i \leq n$. 

ii) $v_i^* = 0$ and $0< v_j^* <1$ for $2\leq i,j \leq n$. In this scenario, $\SA_i$ does not send messages. 
It is equivalent to the situation that $n{-}1$ benign users participate in the competition. 
Thus, the interior point leads to the minimum $\SV(\mathbf{d}^*)$ for fixed $n$.

The interior point results in the minimum $\SV$ given that the top $n$ benign users send messages with positive rates 
at the NE. However, it might not lead to the global minimum, i.e. the minimum for each $n \in[1, N]$. We 
need to compare the minimum $\SV(\mathbf{d}^*)$ for each different $n$. Naturally, the minimum 
$\SV(\mathbf{d}^*)$ is any $n$ is formally given by
\begin{eqnarray}
\SV_{mal} \!\!\!\!\!&\geq&\!\!\!\!\!
\frac{(1{-}n\theta)(1{+}(n{-}1)v_b^{(n)})}{1{+}n\theta} \nonumber\\
\!\!\!\!\!&&\!\!\!\!\!+ (1{+}\frac{n{-}1}{v_b^{(n)}})
(\frac{\theta(1{+}(n{-}1)v_b^{(n)})}{1{+}n\theta})^2\}, \;n=1,{\cdots},N
\label{eq:_theorem5_proof_no10}
\end{eqnarray} 
where $v_b^{(n)}$ is the solution to Eq.\eqref{eq:theorem5_proof_no9} in $(0,1)$.  
Therefore, the lower bound of $\frac{\SV_{mal}}{\SV_{max}}$ is the minimum of 
Eq.\eqref{eq:theorem5_proof_no6.1} and Eq.\eqref{eq:_theorem5_proof_no10}.

When $\theta$ is large, the positive root of Eq.\eqref{eq:theorem5_proof_no9} approaches to 1. 
Then, the inequality \eqref{eq:_theorem5_proof_no10} degenerates to 
\begin{eqnarray}
\SV_{mal} \geq \frac{n}{(1+\theta n)^2}, \quad n=1,{\cdots},N.
\label{eq:_theorem5_proof_no11}
\end{eqnarray} 
We take the derivative of the righthand over $n$ and find that the expression 
$\frac{n}{(1+\theta n)^2}$ is strictly decreasing with $\theta$ and $n$ if $\theta\times n$ is greater than 1. 
Hence, when $\theta$ and $n$ are sufficiently large, Eq.\eqref{eq:_theorem5_proof_no10} is approximated 
by 
\begin{eqnarray}
\SV_{mal} \geq \frac{N}{(1+\theta N)^2}.
\label{eq:_theorem5_proof_no12}
\end{eqnarray} 
The inequalities \eqref{eq:theorem5_proof_no6.1} and \eqref{eq:_theorem5_proof_no12} yield the following approximated lower bound
\begin{eqnarray}
\frac{\SV_{mal}}{\SV_{max}} \geq \min\{\frac{1}{N}, \frac{N}{(1+\theta N)^2}\}. 
\label{eq:_theorem5_proof_no13}
\end{eqnarray}

(4) \textit{\textbf{B4:} (NE Net Utility With/Without Misbehaving User)}

Consider two games, $\mathbf{G_A}$ and $\mathbf{G_B}$: the former excludes the misbehaving user,
the latter considers the possible participation of the misbehaving user. 
Suppose that $n_A$ benign users participate in $\mathbf{G_A}$ at the NE and $n_B$ 
participate in $\mathbf{G_B}$ at the NE. It is easy to prove by contradiction to show 
$n_A \geq n_B$. 

Knowing from the proof of lower bound of \textbf{B3}, 
we have 
\begin{eqnarray}
\SV_{nom} &=& \sum\nolimits_{i=1}^{n_A}v_i - \frac{n_A^2-1}{\sum\nolimits_{i=1}^{n_A}1/v_i}
\label{eq:proof_theorem_SVmalSVnom_no1}
\end{eqnarray}
for the game $\mathbf{G_A}$. For the game $\mathbf{G_B}$, there exist
\begin{eqnarray}
\SV_{mal} = \frac{1{-}n_B\theta}{1{+}n_B\theta} \sum_{i=1}^{n_B}v_i  + (\sum_{i=1}^{n_B}\frac{1}{v_i})\frac{(\theta\sum\nolimits_{i=1}^{n_B}v_i)^2}{(n_B\theta{+}1)^2}
\label{eq:proof_theorem_SVmalSVnom_no2}
\end{eqnarray}
if the misbehaving user participates at the NE, and $\SV_{mal} = \SV_{nom}$ if it does not participate at the NE. 
We then prove this theorem via three steps.

\emph{Step 1: Proving that $\sum\nolimits_{i=1}^{n}v_i - \frac{n^2-1}{\sum\nolimits_{i=1}^{n}1/v_i}$  decreases w.r.t. $n\;\; (2\leq n \leq n_A)$.} 

According to the NE conditions, there must have 
\begin{eqnarray}
v_{n} \geq \frac{n-2}{\sum\nolimits_{j=1}^{n{-}1}1/v_j},\quad \forall 2\leq n \leq n_A. 
\label{eq:proof_theorem_SVmalSVnom_no3}
\end{eqnarray}
Otherwise, the number of benign users participating at the NE will be less than $n_A$. 

Denote by $\Phi(n) = \sum\nolimits_{i=1}^{n}v_i - \frac{n^2-1}{\sum\nolimits_{i=1}^{n}1/v_i}$.
We prove by induction that $\Phi(n) $  decreases w.r.t. $n\;\; (2\leq n \leq n_A)$.

i) For $n=2$, there have $\Phi(1) = 1$ and $\Phi(2) = 1+v_2 - \frac{3v_2}{1+v_2}$. We subtract $\Phi(2)$ from $\Phi(1)$ and obtain
\begin{eqnarray}
\Phi(1) - \Phi(2) = \frac{3}{1+v_2} - v_2 = \frac{1-(1-v_2)^2}{1+v_2} >0 . \nonumber
\end{eqnarray}

ii) we assume $\Phi(n-1) > \Phi(n)$ for $n=k$ \;($2\leq k \leq n_A - 1$). We expand the expression of $\Phi(k)$ and $\Phi(k-1)$ and obtain
\begin{eqnarray}
&&\Phi(k) - \Phi(k{-}1)  \nonumber\\
&=& \sum\nolimits_{i=1}^{k}v_i - \frac{k^2-1}{\sum\nolimits_{i=1}^{k}1/v_i} - 
\sum\nolimits_{i=1}^{k{-}1}v_i + \frac{(k{-}1)^2-1}{\sum\nolimits_{i=1}^{k{-}1}1/v_i} \nonumber\\
&=& v_k - \frac{k^2-1}{\sum\nolimits_{i=1}^{k}1/v_i} + \frac{(k{-}1)^2-1}{\sum\nolimits_{i=1}^{k{-}1}1/v_i}  \nonumber\\
&=& \frac{v_k\sum\nolimits_{i=1}^{k}\frac{1}{v_i}\sum\nolimits_{i=1}^{k{-}1}\frac{1}{v_i}+((k-1)^2-1)\sum\nolimits_{i=1}^{k}\frac{1}{v_i}}{\sum\nolimits_{i=1}^{k}\frac{1}{v_i}\sum\nolimits_{i=1}^{k{-}1}\frac{1}{v_i}} \nonumber\\
&& - \frac{(k^2-1)\sum\nolimits_{i=1}^{k{-}1}\frac{1}{v_i}}{\sum\nolimits_{i=1}^{k}\frac{1}{v_i}\sum\nolimits_{i=1}^{k{-}1}\frac{1}{v_i}}\nonumber\\
&=& \frac{v_k(\sum\nolimits_{i=1}^{k{-}1}\frac{1}{v_i})^2 +(2-2k)\sum\nolimits_{i=1}^{k{-}1}\frac{1}{v_i} +\frac{(k^2-2k)}{v_k}
}{\sum\nolimits_{i=1}^{k}\frac{1}{v_i}\sum\nolimits_{i=1}^{k{-}1}\frac{1}{v_i}} < 0.
\label{eq:proof_theorem_SVmalSVnom_no4}
\end{eqnarray}

iii) For $n=k+1$, we subtract $\Phi(k)$ from $\Phi(k+1)$. There has
\begin{eqnarray}
&&\Phi(k+1) - \Phi(k) \nonumber\\
&=& \frac{v_{k+1}(\sum\nolimits_{i=1}^{k}\frac{1}{v_i})^2 -2k\sum\nolimits_{i=1}^{k}\frac{1}{v_i} +\frac{(k^2-1)}{v_{k+1}}
}{\sum\nolimits_{i=1}^{k{+}1}\frac{1}{v_i}\sum\nolimits_{i=1}^{k}\frac{1}{v_i}}  \nonumber\\
&=& \frac{v_{k+1}(\sum\nolimits_{i=1}^{k{-}1}\frac{1}{v_i})^2 + (2-2k)\sum\nolimits_{i=1}^{k{-}1}\frac{1}{v_i} + \frac{(k^2-2k)}{v_{k{+}1}}}
{\sum\nolimits_{i=1}^{k{+}1}\frac{1}{v_i}\sum\nolimits_{i=1}^{k}\frac{1}{v_i}}. 
\label{eq:proof_theorem_SVmalSVnom_no5}
\end{eqnarray}

The denominator of Eq.\eqref{eq:proof_theorem_SVmalSVnom_no5} can be easily shown to be strictly convex
on $v_{k+1}$. When $v_k$ takes the value $\frac{k-1}{\sum\nolimits_{j=1}^{k}1/v_j}$, there has 
$\Phi(k) - \Phi(k{-}1) =0$. When $v_{k+1}$ takes the value $v_{k}$, the denominator equals to that of 
Eq.\eqref{eq:proof_theorem_SVmalSVnom_no4}, which is less than 0. 
Therefore, we can conclude $\Phi(n) \leq \Phi(n-1)$ for all $2\leq n\leq n_A$.

\emph{Step 2: Approximating the ratio of two net utilities.} 
The ratio $\frac{\SV_{mal}}{\SV_{nom}}$ can be lower bounded by 
\begin{eqnarray}
\frac{\SV_{mal}}{\SV_{nom}} \geq \frac{\frac{1{-}n_B\theta}{1{+}n_B\theta} \sum_{i=1}^{n_B}v_i  {+} (\sum_{i=1}^{n_B}\frac{1}{v_i})\frac{(\theta\sum\nolimits_{i=1}^{n_B}v_i)^2}{(n_B\theta{+}1)^2}}{\sum\nolimits_{i=1}^{n_B}v_i - \frac{n_B^2-1}{\sum\nolimits_{i=1}^{n_B}1/v_i}}
\label{eq:proof_theorem_SVmalSVnom_no6}
\end{eqnarray}
for $n_B\geq 2$ due to the analysis in \emph{Step 1}. For the special case $n_B=1$, the denominator is 1, the maximum achievable
total net utility. Therefore, the lower bound of $\frac{\SV_{mal}}{\SV_{nom}}$ can be expressed as 
Eq.\eqref{eq:proof_theorem_SVmalSVnom_no6} for $n_B\geq 1$. For the sake of convenience, 
we omit the subscript in the variable $n_B$. 

\emph{Step 3: Computing the lower bound of $\frac{\SV_{mal}}{\SV_{nom}}$.} 
Let $\Delta$ be $\sum\nolimits_{i=1}^{n}v_i\sum\nolimits_{i=1}^{n}1/v_i$. 
The lower bound can be simplified as 
\begin{eqnarray}
\frac{\SV_{mal}}{\SV_{nom}} \geq \frac{1}{(1+n\theta)^2}\frac{(1-n^2\theta^2)\Delta + \theta^2\Delta^2}{\Delta - (n^2-1)}.
\label{eq:proof_theorem_SVmalSVnom_no7}
\end{eqnarray}
We differentiate the right-hand expression of Eq.\eqref{eq:proof_theorem_SVmalSVnom_no7} 
over $\Delta$ and find that the derivative is positive. 
This means that the minimum of the right-hand expression is obtained when $\Delta$ is minimized. 
Since the minimum $\Delta$ is $n^2$ obtained at $v_i=1$ for all $1\leq i \leq n$, the lower bound of the ratio $\frac{\SV_{mal}}{\SV_{nom}}$ is 
\begin{eqnarray}
\frac{\SV_{mal}}{\SV_{nom}} \geq \frac{n^2}{(1+n\theta)^2}.
\label{eq:proof_theorem_SVmalSVnom_no8}
\end{eqnarray}
The expression $\frac{n^2}{(1+n\theta)^2}$ is strictly increasing w.r.t. $n$. We finally conclude the worst case of the ratio 
as 
\begin{eqnarray}
\frac{\SV_{mal}}{\SV_{nom}} \geq \frac{1}{(1+\theta)^2}.
\label{eq:proof_theorem_SVmalSVnom_no9}
\end{eqnarray}

(5) \textit{\textbf{B5:} (Operator's Revenue over Maximum)}

Note that $\SW_{max}$ is the maximum revenue obtained by the network operator excluding the player $\SA_0$. 
The NE conditions yield 
\begin{eqnarray}
\SW_{mal} =  \sum\nolimits_{i=0}^nx_i^* = \frac{\theta\sum\nolimits_{i=1}^{n}v_i}{1+n\theta}
\label{eq:SW_malvsopt_no1}
\end{eqnarray} 
when the top $n$ benign users and the misbehaving user participate at the NE;  
they also give rise to 
\begin{eqnarray}
\SW_{mal} =  \sum\nolimits_{i=1}^nx_i^* = \frac{n-1}{\sum\nolimits_{i{=}1}^{n}1/v_i} 
\label{eq:SW_malvsopt_no2}
\end{eqnarray} 
when the misbehaving user does not participate. 
It is clear to observe that $\SW_{mal}$ is an increasing function of any $v_i$ ($2\leq i\leq n$). 
When the misbehaving user is excluded, the revenue of the network operator is given by 
\begin{eqnarray}
\SW_{max} = \max_{\{v_i\}_{i{=}2}^N} \frac{n-1}{\sum\nolimits_{i{=}1}^{n}1/v_i} 
\label{eq:SW_malvsopt_no3}
\end{eqnarray} 
where $n$ is determined by $\{v_i\}_{i{=}2}^N$. Similarly, Eq.\eqref{eq:SW_malvsopt_no3} is also an increasing function of any $v_i$ ($2\leq i\leq n$). By enumerating all the possible $n$, we can see 
\begin{eqnarray}
\SW_{max} = \frac{N-1}{N} .
\end{eqnarray}

\emph{Proving the lower bound}. Eq.\eqref{eq:SW_malvsopt_no1} and \eqref{eq:SW_malvsopt_no2} 
both show that $\SW_{mal}$ is the increasing function of any $v_i$ ($2\leq i\leq n$). 
Thus, when the minimum $\SW_{mal}$ is obtained, all the benign users from $\SA_2$ to 
$\SA_{n}$ have the same lowest valuation denoted by $\underline{v}$. 
However, we need to find $\underline{v}$ for each case, $x_0^* > 0$ and $x_0^* = 0$.

Case i): $x_0^* > 0$. 
The NE conditions yield
\begin{eqnarray}
&&\frac{1}{\sum\nolimits_{i=0}^nx_i^* } - \frac{x_1^*}{(\sum\nolimits_{i=0}^nx_i^*)^2} = 1 , 
\label{eq:SW_malvsopt_no4}\\
&&\frac{\underline{v}}{\sum\nolimits_{i=0}^nx_i^* } - \frac{\underline{v}x_j^*}{(\sum\nolimits_{i=0}^nx_i^*)^2} = 1 , \quad \forall 2\leq j \leq n,
\label{eq:SW_malvsopt_no5}\\
&&\frac{x_1^*+(n-1)\underline{v}x_j^*}{(\sum\nolimits_{i=0}^nx_i^*)^2 }  = \frac{1}{\theta} .
\label{eq:SW_malvsopt_no6}
\end{eqnarray} 
When the benign users from $\SA_2$ to $\SA_n$ are at the boundary of participation, we obtain $\underline{v} = \sum\nolimits_{i=0}^nx_i^*$
from Eq.\eqref{eq:SW_malvsopt_no5}, and $x_0^*+x_1^* = \frac{\theta}{1+\theta}$ 
from Eq.\eqref{eq:SW_malvsopt_no4} and Eq.\eqref{eq:SW_malvsopt_no6} 
by letting $x_j^*$ be infinitely small. 
Then, the minimum valuation $\underline{v}$ is given by
\begin{eqnarray}
\underline{v} = \frac{\theta}{1+\theta}.
\label{eq:SW_malvsopt_no9}
\end{eqnarray} 
The minimum revenue of the network operator is 
\begin{eqnarray}
\SW_{mal} \geq \frac{\theta}{1+\theta}.
\label{eq:SW_malvsopt_no9_2}
\end{eqnarray} 

Case ii): $x_0^* = 0$. The minimum valuation $\underline{v}$ should be above a certain threshold such 
that the misbehaving user will not participate in the competition. 
According to the NE conditions, there has 
\begin{eqnarray}
\frac{x_1^*+(n-1)\underline{v}x_j^*}{(\sum\nolimits_{i=0}^nx_i^*)^2 }  \leq  \frac{1}{\theta} .
\label{eq:SW_malvsopt_no10}
\end{eqnarray} 
Summing up Eqs.\eqref{eq:SW_malvsopt_no4}, \eqref{eq:SW_malvsopt_no5} and \eqref{eq:SW_malvsopt_no10} 
together, we obtain
\begin{eqnarray}
\sum\nolimits_{i=0}^nx_i^* = \sum\nolimits_{i=1}^nx_i^* = \frac{n-1}{(1+(n-1)/\underline{v})} \leq \frac{\theta(1+(n-1)\underline{v})}{1+n\theta}
\label{eq:SW_malvsopt_no11}
\end{eqnarray} 
due to $x_0^* = 0$. The equality holds when $\underline{v}$ reaches the stage that the misbehaving user 
is at the boundary of participation. Eq.\eqref{eq:SW_malvsopt_no11} gives rise to the 
following equation to solve $\underline{v}$,
\begin{eqnarray}
\underline{v}^2 + (\frac{1}{n-1}-(1+\frac{1}{\theta}))\underline{v} + 1 = 0.
\label{eq:SW_malvsopt_no12}
\end{eqnarray} 
Here, $\theta$ is no larger than $\frac{n-1}{n}$. Otherwise, the misbehaving user will always 
participate in the competition. We next examine whether there exists a feasible solution 
to $\underline{v}$ in Eq.\eqref{eq:SW_malvsopt_no12}. 
The expression $\frac{1}{n-1}-(1+\frac{1}{\theta})$ is always negative with $\theta \in [0,\frac{n-1}{n}]$. 
This means that there are two positive roots to Eq.\eqref{eq:SW_malvsopt_no12}. The product of these two 
roots is 1, which indicates that one root is in the range $(0, 1)$ and the other is greater than 1.  
Thus, we solve $\underline{v}$ by
\begin{eqnarray}
\underline{v} = \frac{(1+\frac{1}{\theta})-\frac{1}{n-1} - \sqrt{((1+\frac{1}{\theta})-\frac{1}{n-1})^2-4}}{2} .
\label{eq:SW_malvsopt_no13}
\end{eqnarray} 
When the valuations of $\SA_2$ to $\SA_n$ take $\underline{v}$ in Eq.\eqref{eq:SW_malvsopt_no13}, 
both Eq.\eqref{eq:SW_malvsopt_no1} and Eq.\eqref{eq:SW_malvsopt_no2} 
yield the same revenue to the network operator at this boundary condition. 

So far, we have computed two possible $\underline{v}$ for two cases $x_0^* = 0$ and $x_0^*>0$ separately. 
We denote by $\underline{v}_A$ the minimum valuation in Eq.\eqref{eq:SW_malvsopt_no13} and by $\underline{v}_B$ the minimum valuation in Eq.\eqref{eq:SW_malvsopt_no9}. 
The final question is whether $\underline{v}_A$ or $\underline{v}_B$ 
results in the minimum revenue of the network operator. We subtract $\underline{v}_B$ from $\underline{v}_A$ and obtain
\begin{eqnarray}
\!\!\!\! \!\!\!\! \!\!\!\!&\!\!\!\! \!\!\!\! \!\!\!\!&\!\!\!\! \!\!\!\! \!\!\!\!\underline{v}_A - \underline{v}_B \nonumber\\
\!\!\!\! \!\!\!\! \!\!\!\!&\!\!\!\! =&\!\!\! \frac{(1+\frac{1}{\theta})-\frac{1}{n-1} - \sqrt{((1+\frac{1}{\theta})-\frac{1}{n-1})^2-4}}{2} - \frac{\theta}{1+\theta} \nonumber\\
\!\!\!\! \!\!\!\! \!\!\!\!&\!\!\!\! =&\!\!\!  \frac{1}{2}\Big[\big((1\!+\!\frac{1}{\theta})\!-\!\frac{1}{n\!-\!1} \!-\! \frac{2\theta}{1\!+\!\theta}\big) 
\!-\! \sqrt{((1\!+\!\frac{1}{\theta})-\frac{1}{n\!-\!1})^2\!-\!4}\Big].
\label{eq:SW_malvsopt_no14}
\end{eqnarray} 
Since $\frac{2\theta}{1+\theta}$ is less than 2, it is easy to validate that $\underline{v}_A$ is always 
greater than $\underline{v}_B$.
Our previous analysis has shown that the revenue of the network operator is an 
increasing function of $\underline{v}$ 
no matter whether the misbehaving user participate or not. 
Therefore, the minimum revenue should be obtained in the case $x_0^* > 0$. 
To summarize, we have the following lower bound 
\begin{eqnarray}
\frac{\SW_{mal}}{\SW_{max}} \geq \frac{\theta N}{(1+\theta)(N-1)}. 
\label{eq:SW_malvsopt_no15}
\end{eqnarray}

(6) \textit{\textbf{B6:} (Operator's Revenue With/Without Misbehaving User)} 

Note that $\SW_{mal}$ is the same as $\SW_{nom}$ if the misbehaving user does not participate at the NE. 
Hence, we only consider the scenario with the participation of the misbehaving user. 
Similar to the proceeding proofs, we denote $\mathbf{G_A}$ as the game excluding the misbehaving user, and 
$\mathbf{G_B}$ as that with the possible participation of the misbehaving user. 
Suppose that $n_A$ benign users send messages with positive rates at the NE of $\mathbf{G_A}$ and 
$n_B$ benign users do so at the NE of $\mathbf{G_B}$. 
Then, according to Eqs.\eqref{eq:SW_malvsopt_no1} and \eqref{eq:SW_malvsopt_no2}, 
there exists $n_A \geq n_B$. The ratio $\frac{\SW_{mal}}{\SW_{nom}}$ satisfies 
\begin{eqnarray}
\frac{\SW_{mal}}{\SW_{nom}} = \frac{\theta \sum\nolimits_{i=1}^{n_B}v_i\sum\nolimits_{i=1}^{n_A}1/v_i}{(1+n_B\theta)(n_A-1)}
\label{eq:SW_malvsnom_no1}
\end{eqnarray} 
According to the inequality \eqref{eq:SU_malvsnom_no5}, this ratio has the following bound
\begin{eqnarray}
\frac{\SW_{mal}}{\SW_{nom}} \geq  \frac{\theta \sum\nolimits_{i=1}^{n_A}v_i\sum\nolimits_{i=1}^{n_A}1/v_i}{(1+n_A\theta)(n_A-1)}
\label{eq:SW_malvsnom_no2}
\end{eqnarray} 
so that we only need to analyze the lower bound with only the benign users. 
In what follows, the subscripts on the variable $n$ are removed for simplicity. 
To obtain the lower bound of $\frac{\SW_{mal}}{\SW_{nom}}$, we need to find the minimum for 
the expression $\sum\nolimits_{i=1}^{n}v_i\sum\nolimits_{i=1}^{n}1/v_i$. In the step 3 of proof of Theorem 5, 
the above expression is shown to be a decreasing function of $v_n$. Hence, 
it is intuitive to see that $v_i (\forall i\geq 2)$ should be as large as possible at the 
right-hand of Eq.\eqref{eq:SW_malvsnom_no2}. Hence, we have 
\begin{eqnarray}
\frac{\SW_{mal}}{\SW_{nom}} \geq  \frac{\theta N^2}{(1+\theta N)(N-1)}.
\label{eq:SW_malvsnom_no3}
\end{eqnarray} 
For the given $\theta$, $\mathbf{v}$ should be chosen to allow the participation of the misbehaving user. 
Therefore, according to the NE conditions, the maximum ratio is obtained by $\{v_i\}_{i=2}^{n}$ chosen from
\begin{eqnarray}
\frac{\theta \sum_{i=1}^{n}v_i}{1+n\theta} = \frac{n-1}{\sum\nolimits_{i=1}^{n} 1/v_i} .
\label{eq:SW_malvsnom_no3}
\end{eqnarray} 
The above equality gives rise to 
\begin{eqnarray}
\sum\nolimits_{i=1}^{n}v_i\sum \frac{1}{v_i} \geq \frac{1}{\theta}(n-1)(1+n\theta). 
\label{eq:SW_malvsnom_no4}
\end{eqnarray} 
Submitting inequality \eqref{eq:SW_malvsnom_no4} to \eqref{eq:SW_malvsnom_no2}, we obtain the lower bound as
\begin{eqnarray}
\frac{\SW_{mal}}{\SW_{nom}} \geq \max\{1, \frac{\theta N^2}{(1+\theta N)(N-1)}\}.
\label{eq:SW_malvsnom_no5}
\end{eqnarray} 
\done


\subsection*{Proof of Theorem \ref{theorem:all_upperbounds_linear}}

(1) \textit{\textbf{B1:} (NE Utility over Maximum)}

Similarly, we consider two cases separately, $x_0^* = 0$ and $x_0^* > 0$. Note that the utility 
functions of all the benign users are linear. 

\noindent \textit{Case 1: $x_0^*=0$.} It is very direct to validate that the
maximum total utility is 1. For instance, when $v_i$ is equal to $1$ for 
all $1\leq i\leq N$, the total utility is maximized. However, the set of
valuations $\{v_i\}$ that lead to this maximum are not unique.
We hereby want to find the condition to enforce $x_0^*=0$. 

Suppose that $n$ benign users participate in the competition at the NE. For an arbitrary set of 
valuations $\{v_i\}_{i=2}^n$, the NE conditions yield 
\begin{eqnarray}
\theta \leq \sum\nolimits_{i=1}^{n}x_i^*
\label{eq:theorem6_proof_no1}
\end{eqnarray}
for $x_0^* = 0$. It is easy to validate from our preceding proof that $x_i^*$ increases with $v_i$. At the 
same time, when more benign users participate at the NE, the total rate of sending messages increases 
accordingly. Hence, $x_0^*=0$ is no longer true when there has $\theta \leq \frac{N-1}{N}$.

\noindent \textit{Case 2: $x_0^*>0$.} We suppose $x_i^* > 0$ for 
$2\leq i \leq n$ and  $x_i^* = 0$ for $n{+}1\leq i\leq N$. According to
Eq. \eqref{eq:proof_totutility_withA0}, the total utility of benign users is given by 
$\SU(\mathbf{d}^*) = \frac{\sum\nolimits_{j=1}^{n}v_j}{n\theta+1}$.
The maximum total utility is achieved at the point $v_i=1$ for all $1\leq i,j\leq n$, 
that is, $\SU(\mathbf{d}^*) \leq \frac{n}{nd+1}$. As $n$ increases, this upper bound
increases accordingly. Then, there has $\SU(\mathbf{d}^*) \leq \frac{N}{Nd+1}$ for any $n$.
When $\theta$ is greater than $\frac{N-1}{N}$, $x_0^*$ is always positive. 
The upper bound of the maximum total utility is given by
\begin{eqnarray}
\SU_{mal} \leq \frac{N}{1+\theta N}.
\end{eqnarray}
Combining the analyses in two cases together, we prove this theorem. 

(2) \textit{\textbf{B2:} (Approximate Upper Bound of NE Utility With/Without Misbehaving User)} 


Consider two games, $\mathbf{G_A}$ and $\mathbf{G_B}$: the former excludes the misbehaving user,
the latter considers the possible participation of the misbehaving user. 
Suppose that $n_A$ benign users participate in $\mathbf{G_A}$ at the NE and $n_B$ 
participate in $\mathbf{G_B}$ at the NE.

When the misbehaving user does not participate at the NE of $\mathbf{G_B}$, both NEs are the same 
so that the ratio $\frac{\SU_{mal}}{\SU_{nom}}$ is 1. When $\theta$ is greater than $\frac{N{-}1}{N}$, the misbehaving user will participate for sure. In this scenario, it is difficult to compute the 
upper bound of $\frac{\SU_{mal}}{\SU_{nom}}$. The reason is that $\frac{\SU_{mal}}{\SU_{nom}}$ with
$n_A=n_B$ does not necessarily constitute an upper bound for each $n$. Hence, we only aim to find an 
approximated upper bound for $\frac{\SU_{mal}}{\SU_{nom}}$.

i). We first consider the case that $\theta$ is slightly larger than $\frac{N-1}{N}$. 
When $\theta$ is less than $\frac{N-1}{N}$, the maximum total utility is obtained 
when the valuations of the participating users are 1 uniformly. 
When $\theta$ is slightly larger than $\frac{N-1}{N}$, the number of the benign users that 
participate in the game $\mathbf{G_B}$ will not change. Because $\SU_{mal}$ 
is an increasing function of the valuations, the maximum $\SU_{mal}$ is 
obtained by $\frac{n}{1+n\theta}$ where $n$ is the number of the benign users that participate. 
Then, the upper bound of the ratio $\frac{\SU_{mal}}{\SU_{nom}}$ is approximated by
\begin{eqnarray}
\frac{\SU_{mal}}{\SU_{nom}} \leq \frac{\sum\nolimits_{i=1}^{n}v_i\sum\nolimits_{i=1}^{n}1/v_i}
{\sum\nolimits_{i=1}^{n}v_i\sum\nolimits_{i=1}^{n}1/v_i - n(n-1)} \cdot \frac{1}{1+n\theta}.
\end{eqnarray} 
Following the proof of lower bound of \textbf{B2}, 
the minimum of the expression 
$\sum\nolimits_{i=1}^{n}v_i\sum\nolimits_{i=1}^{n}1/v_i$ is $n^2$. The lower bound of 
$\frac{\SU_{mal}}{\SU_{nom}}$ can be further simplified as
\begin{eqnarray}
\frac{\SU_{mal}}{\SU_{nom}} \leq \frac{n}{1+\theta n} \leq \frac{N}{1+N\theta}.
\end{eqnarray}

ii) We next consider the case that $\theta$ is large enough. 
In the game $\mathbf{G_B}$, $n_B$ might be larger if the valuations of the benign users 
increase. The highest improvement takes place when $n_B$ increases from 1 to 2. 
However, $\SW_{mal}$ only increases from $\frac{1}{1+\theta}$ to $\frac{2}{2+\theta}$, which is 
very small for some large $\theta$. 
In the game $\mathbf{G_A}$, the minimum $\SU_{nom}$ is given by Eq.\eqref{eq:theorem_lowerbounds_no1} 
in Theorem \ref{theorem:all_lowerbounds_linear}. The minimum is achieved when all the benign users 
participate in the competition at the NE. The valuations of $\SA_2$ to $\SA_N$ are $\sqrt{N(N-1)-(N-1)}$ 
uniformly. When $v_2$ increases to $1$, $\SU_{nom}$ improves significantly. Hence, when $\theta$
is large enough, the approximated upper bound of $\frac{\SU_{mal}}{\SU_{nom}}$ takes place 
at $n_B=1$ and $n_A=N$ where the minimum $\SU_{nom}$ is reached. 
Then, the approximated upper bound is given by
\begin{eqnarray}
\frac{\SU_{mal}}{\SU_{nom}} \leq \frac{1}{(1+\theta)(1-(N-1)(\sqrt{N}-\sqrt{N-1})^2)}.
\end{eqnarray} 

Combing the above two cases together, we obtain the approximated upper bound for the ratio 
$\frac{\SU_{mal}}{\SU_{nom}}$ as
\begin{eqnarray}
\frac{\SU_{mal}}{\SU_{nom}} &\!\!\!\leq\!\!\!& \max\{\frac{N}{1+N\theta}, \nonumber\\
&\!\!\!\!\!\!&\frac{1}{(1+\theta)(1-(N-1)(\sqrt{N}-\sqrt{N-1})^2)}\}.
\end{eqnarray} 
\done

(3) \textit{\textbf{B3:} (NE Net Utility Over Maximum)}


We suppose that $x_i^*>0$ for $1\leq i\leq n$ and $x_i^*=0$ for 
$n{+}1\leq i\leq N$ at the NE. The optimal total net utility is found 
for each $n$ in the first step. We then compare the optimal total net utilities
among all the possible $n$. Two cases are considered, 
$x_0^*=0$ or  $x_0^*>0$ (i.e. whether the misbehaving user sends messages to the timeline or not).

\noindent \textit{Case 1: $x_0^*=0$.} The total net utility of benign users is given by
\begin{eqnarray}
\SV(\mathbf{d}^*) = \sum\nolimits_{i=1}^{n}v_i - \frac{n^2-1}{\sum\nolimits_{i=1}^{n}1/v_i}.
\label{eq:theorem7_proof_no1}
\end{eqnarray}
According to the analysis in the proof of lower bound of \textbf{B1}, 
$\SV(\mathbf{d}^*)$
is a strictly convex function so that the maximum $\SV(\mathbf{d}^*)$ is obtain at the boundary
of the feasible region, i.e. $v_i=1$ for all $1\leq i\leq n$. In other words,
there has $\max\;\SV(\mathbf{d}^*) = \frac{1}{n}$. This maximum is achieved under the
condition $\theta<\frac{n-1}{n}$ such that the misbehaving user $\SA_0$ does not participate. 

\noindent \textit{Case 2: $x_0^*>0$.} The total net utility of benign users is given in Eq.\eqref{eq:theorem5_proof_no7}. 
The proof of lower bound of \textbf{B3} 
manifests that there exists a unique local minimum. The maximum
total net utility should be obtained at the boundary. Here, we only consider the right boundary, i.e. $v_i=1$ for $2\leq i\leq n$,
because the left boundary corresponds to the participation of less than $n$ benign users. Then, the maximum total net utility is obtained by
\begin{eqnarray}
\SV(\mathbf{d}^*) = \frac{n}{(nd+1)^2}. \nonumber
\label{eq:theorem7_proof_no2}
\end{eqnarray}
In order to guarantee $x_0^*>0$ at the NE, the willingness factor $\theta$ must satisfy $\theta\geq \frac{n-1}{n}$.

We then proceed to find the maximum $\SV(\mathbf{d}^*)$ for all $n\in[1, N]$ 
that satisfies the corresponding condition $\theta \geq \frac{n-1}{n}$. Let $n_1$ and $n_2$
be two different integers in the set $[1,N]$. We compare the total net utilities at these two different scenarios. 
\begin{eqnarray}
\SV(\mathbf{d}^*)|_{n_1} - \SV(\mathbf{d}^*)|_{n_2} = \frac{(n_1-n_2)(1-n_1n_2\theta^2)}{(1+n_1\theta)^2(1+n_2\theta)^2}. \nonumber
\label{eq:theorem7_proof_no3}
\end{eqnarray}
By endowing $n_1$ and $n_2$ different values, we obtain

- $\SV(\mathbf{d}^*)|_{1} < \SV(\mathbf{d}^*)|_{2}$ if $\frac{1}{2}<\theta < \frac{\sqrt{2}}{2}$;

- $\SV(\mathbf{d}^*)|_{1} <\SV(\mathbf{d}^*)|_{3}$ is not true (i.e. $\theta\in (0,\frac{\sqrt{3}}{3})\cap(\frac{2}{3},\infty)= \varnothing$);

- $\SV(\mathbf{d}^*)|_{1} < \SV(\mathbf{d}^*)|_{4}$ is not true (i.e. $\theta \in (0,\frac{1}{2})\cap(\frac{3}{4},\infty)= \varnothing$);

- $\SV(\mathbf{d}^*)|_{2} < \SV(\mathbf{d}^*)|_{3}$ is not true (i.e. $d\in (0,\frac{\sqrt{6}}{6})\cap(\frac{2}{3},\infty)= \varnothing$).

This is to say, for any $\theta$, the maximum $\SV(\mathbf{d}^*)$ is either $\SV(\mathbf{d}^*)|_{1}$ or $\SV(\mathbf{d}^*)|_{2}$.

We next merge the analyses of the cases $x_0^* =0$ and $x_0^* > 0$. 
When $\theta$ is in the range $(\frac{n-2}{n-1}, \frac{n-1}{n})$, the optimal
total net utility is $\frac{1}{n}$ if $x_0^*=0$ and $n\geq 2$. The optimal 
total net utility is $\frac{n{-}1}{(1+(n{-}1)\theta)^2}$ if $x_0^*>0$. 
When $n\geq 3$ and $\theta \in
(\frac{n-2}{n-1}, \frac{n-1}{n})$, the following expression always holds
\begin{eqnarray}
\frac{n{-}1}{(1+(n{-}1)\theta)^2} > \frac{1}{n}, \quad \forall \; n\geq 3.  \nonumber
\label{eq:theorem7_proof_no4}
\end{eqnarray}
The participation of $\SA_0$ at the NE always generates a better upper bound of the 
total net utility of benign users than that in the absence of $\SA_0$.
Therefore, we only need to compare three outcomes, $\frac{1}{(1+\theta)^2}$
for $\theta>0$, $\frac{2}{(1+2\theta)^2}$
for $\theta>\frac{1}{2}$, and $\frac{1}{2}$ for $0<\theta\leq \frac{1}{2}$.

We eventually summarize our results as follows
\begin{eqnarray}
\frac{\SV_{mal}}{\SV_{max}} \leq \left\{\begin{matrix}
\frac{1}{(1+\theta)^2} && \textrm{ if } \theta \leq\sqrt{2}{-}1 \\
\frac{1}{2} && \textrm{ if } \sqrt{2}{-}1 < \theta \leq\frac{1}{2} \\
\frac{2}{(1+2\theta)^2}, && \textrm{ if } \; \frac{1}{2}<\theta\leq\frac{\sqrt{2}}{2}\\
\frac{1}{(1+\theta)^2}, && \textrm{ if } \; \theta >\frac{\sqrt{2}}{2}
\end{matrix}\right..
\end{eqnarray} 
\done

(4) \textit{\textbf{B5:} (Operator's Revenue over Optimality)} 
    

\emph{Proving the upper bound}. If the misbehaving user does not participate at the NE, 
the maximum achievable revenue of the network operator is the same as $\SW_{max}$. 
On the contrary, if the misbehaving user participates at the NE, the maximum $\SW_{mal}$ is obtained by
\begin{eqnarray}
\SW_{mal} \geq \frac{\theta N}{(1+\theta)(N-1)}. 
\label{eq:SW_malvsopt_no16}
\end{eqnarray} 
Then, the upper bound is 
\begin{eqnarray}
\frac{\SW_{mal}}{\SW_{max}} \leq \max\{1, \frac{\theta N^2}{(1+\theta N)(N-1)}\}. 
\label{eq:SW_malvsopt_no17}
\end{eqnarray} 
\done

(5) \textit{\textbf{B6:} (Operator's Revenue With/Without Misbehaving User)} 


The upper bound can be solved directly. Consider a scenario with $v_2=\epsilon$ and $v_i = 0$ for all $i\geq 3$. 
The total revenue of the network operator is small enough with the asymptotic bound $\SW_{nom} = 0$. 
For any given $\theta$ and the participation of the misbehaving user at the NE, the total revenue is finite. 
Therefore, the ratio has the property $\frac{\SW_{mal}}{\SW_{nom}} \propto \infty$ which means that it is unbounded. \done

\subsection*{Proof of Theorem \ref{theorem:no3_lowerboundscondition}}

We will show that the worst case of general utility functions occurs with linear utility functions of benign users. 
	i) \emph{The worst case of \textbf{B1}}. Johari and Tsitsiklis \cite{johari} proved that the
	worse case of \textbf{B1} occurred in the case of linear utility functions of benign users. 
	Although there exists a misbehaving user in our problem, the same proof can be reused without any modification. 

ii) \emph{The worst case of \textbf{B3}}. 
Consider an arbitrary strategy $\bar{\bx}$ and the social optimal strategy $\bx^{S}$ that yield the 
corresponding allocations, 
$\mathbf{\bar{d}}=\{\bar{d}_1, \bar{d}_2,\cdots, \bar{d}_N\}$ and $\mathbf{d^S}=\{d_1^S, d_2^S,\cdots, d_N^S\}$. 
There exist $\bar{x}_i \geq 0$, $\bar{d}_i \geq 0$, $x_i^S \geq 0$, $d_i^S\geq 0$ and $U_{\SA_i}(0) = 0$ for all $i\geq 1$. 

The concavity of $U_{\SA_i}(\bar{d}_i)$ leads to
\begin{eqnarray}
U_{\SA_i}(\bar{d}_i) + U_{\SA_i}'(\bar{d}_i)(d_i^S - \bar{d}_i) \geq U_{\SA_i}(\bar{d}_i^S) , 
\quad i=1{, \cdots, }N.
\end{eqnarray}
Therefore, we obtain a series of inequalities
\begin{eqnarray}
\frac{\SV(\bar{\bx})}{\SV(\bx^{S})} \!\!\!\!\!&=&\!\!\!\! \frac{\sum\nolimits_{i=1}^N(U_{\SA_i}(\bar{d}_i) 
	- \bar{x}_i)}{\sum\nolimits_{i=1}^N(U_{\SA_i}(\bar{d}_i^S) {-} \bar{x}_i^S)} \nonumber\\
\!\!\!\!\!&\geq&\!\!\!\! \frac{\sum\nolimits_{i=1}^N(U_{\SA_i}(\bar{d}_i) 
	{-} \bar{x}_i)}{\sum\nolimits_{i=1}^NU_{\SA_i}(\bar{d}_i^S)} \nonumber\\
&\!\!\!\!\!\!\!\!\!\!\!\!\!\!\!\!\!\!\!\!\!\!\!\!\!\!\!\!\!\!\!\!\!\!\!\!\!
\geq\!\!\!\!\!\!\!\!\!\!&\!\!\!\!\!\!\!\!\!\!\!\!\!\!\!\!\!\!\!\! \frac{\sum\nolimits_{i=1}^N\big(U_{\SA_i}\!(\bar{d}_i){-}U_{\SA_i}'\!(\bar{d}_i)\bar{d}_i\big) {+} \sum\nolimits_{i=1}^N\big(U_{\SA_i}'\!(\bar{d}_i)\bar{d}_i  {-} \bar{x}_i\big)}
{\sum\nolimits_{i=1}^N\big(U_{\SA_i}\!(\bar{d}_i){-}U_{\SA_i}'\!(\bar{d}_i)\bar{d}_i\big) {+} \sum\nolimits_{i=1}^NU_{\SA_i}'\!(\bar{d}_i)d_i^S}. 
\label{eq:proof_theorem3_no2}
\end{eqnarray}
We next derive two inequalities to simplify the above expression. Because $d_i^S$ is a fraction 
with $\sum\nolimits_{i=1}^{N}d_i^S\leq 1$, the following inequality holds, 
$\sum\nolimits_{i=1}^NU_{\SA_i}'\!(\bar{d}_i)d_i^S \leq \max_iU_{\SA_i}'\!(\bar{d}_i)$.
Since $U_{\SA_i}(\bar{d}_i)$ is concave and strictly increasing with $U_{\SA_i}(0) = 0$, 
the expression 
$U_{\SA_i}\!(\bar{d}_i){-}U_{\SA_i}'\!(\bar{d}_i)\bar{d}_i$ is nonnegative. 
Therefore, Eq.\eqref{eq:proof_theorem3_no2} satisfies
\begin{eqnarray}
&&\!\!\!\!\!\!\!\!\!\!\!\!\!\!\!\!\frac{\SV(\bar{\bx})}{\SV(\bx^{S})} \geq \frac{\sum\nolimits_{i=1}^N\big(U_{\SA_i}'\!(\bar{d}_i)\bar{d}_i  {-} \bar{x}_i\big)}{\max_i U_{\SA_i}'\!(\bar{d}_i)}. 
\label{eq:proof_theorem3_no3}
\end{eqnarray}
Let $\bx^*$ be the unique NE strategy, and $\mathbf{d}^*$ be the vector 
of allocations at this NE. 
We define a new class of linear utility functions as 
\begin{eqnarray}
\bar{U}_i(d_i) = U_i'(d_i^*)d_i, \quad \forall 1\leq i \leq N.
\end{eqnarray}
If we let $\bar{\mathbf{x}} = \mathbf{x}^*$ 
(also $\bar{\mathbf{d}} = \mathbf{d}^*$ accordingly), the numerator in the right-hand of 
Eq.\eqref{eq:proof_theorem3_no3} is the NE, and the denominator is the social optimal 
total net utility of benign users and also the social optimal total utility of benign users. 
Therefore, we can see that the worst-case ratio $\frac{\SV_{mal}}{\SV_{max}}$ occurs in the case of 
linear utility functions. 

	
	\begin{figure}[!h]
		\centering
		\includegraphics[width=2.8in]{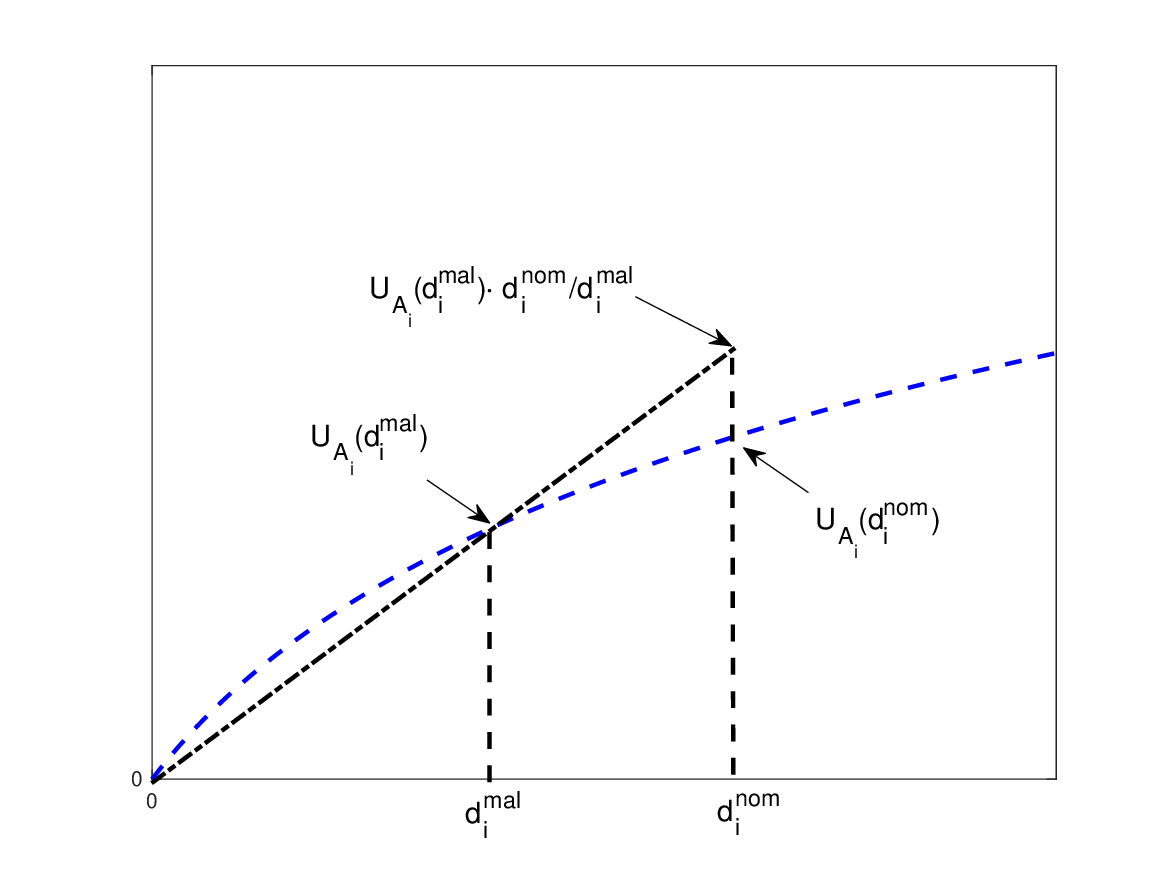}
		\caption{An illustration of the worst case condition of \textbf{B2}.}
		\label{fig:proof_lastone}
	\end{figure}

	iii) \emph{The worst case of \textbf{B2} and \textbf{B4}}. 
	Suppose that the allocations at the NEs of \textbf{MAL} and \textbf{NOM} 
	are denoted by $\mathbf{d}_{mal}^*$$=$$\{d_{1,mal}^*, d_{2,mal}^*,\cdots, d_{N,mal}^*\}$ and 
	$\mathbf{d}_{nom}^*$$=$$\{d_{1,nom}^*, d_{2,nom}^*,\cdots, d_{N,nom}^*\}$.
	According to the KKT conditions Eq.\eqref{eq:onetarget_opt1} and 
	\eqref{eq:onetarget_opt2}, it is easy to conclude $d_{i,nom}^* \geq d_{i,mal}^*$ for all $i\geq 1$. 
	
	Before delivering the proof, we use a simple illustration to explain why the worst case occurs with linear utility functions. 
	Figure \ref{fig:proof_lastone} shows a strictly concave and strictly increasing utility function. 
	One can see that there has
	\begin{eqnarray}
	\frac{U_{\SA_i}(d_i^{mal})}{U_{\SA_i}(d_i^{nom})} > \frac{U_{\SA_i}(d_i^{mal})}{U_{\SA_i}(d_i^{mal})\cdot d_i^{nom}/d_i^{mal}} 
	= \frac{d_i^{mal}}{d_i^{nom}}. \nonumber
	\end{eqnarray}
	Formally, the following inequality holds,
	\begin{eqnarray}
	\frac{U_{\SA_i}(d_i^{mal})}{U_{\SA_i}(d_i^{nom})} \geq \frac{U_{\SA_i}(d_i^{mal})}{U_{\SA_i}(d_i^{mal}) + U'_{\SA_i}(d_i^{mal})(d_i^{nom}-d_i^{mal})} \nonumber
	\end{eqnarray}
	for $0<d_i^{mal}<d_i^{nom}$. The equality holds only upon $U'_{\SA_i}(d_i^{mal}) =0$, that is, the utility function of $\SA_i$ is linear to $d_i$.  Combing the utility functions of all the benign users, we observe that the worse case of \textbf{B2} happens 
	when all the benign users have linear utility functions. Following the same approach, we can also prove the worst case condition 
	of \textbf{B4}. \done

\end{document}